\documentclass[reprint,aps,pra,showpacs,amsmath,amssymb,twocolumn,superscriptaddress]{revtex4-2}
\usepackage[dvips]{graphicx}
\usepackage{amsmath,amssymb,amsthm,mathrsfs,amsfonts,dsfont}
\usepackage{subfigure, epsfig}
\usepackage{braket}
\usepackage{bm}
\usepackage[T1]{fontenc}
\usepackage{enumerate}
\usepackage{color}
\usepackage{qcircuit}
\usepackage{graphicx}
\usepackage{algorithm}
\usepackage{algpseudocode}
\usepackage{algpseudocode}
\usepackage{hyperref}
\hypersetup{colorlinks=true, linkcolor=blue, citecolor=blue, urlcolor=black }
\usepackage{comment}

\usepackage{dcolumn}
\usepackage{bm}
\usepackage{tikz}
\usepackage{ulem}
\usepackage{graphicx}

\usepackage{varioref}
\labelformat{equation}{(#1)}

\labelformat{section}{#1}

\labelformat{figure}{#1}

\labelformat{proposition}{#1}

\labelformat{lemma}{#1}

\labelformat{theorem}{#1}

\labelformat{observation}{#1}

\labelformat{definition}{#1}

\labelformat{corollary}{#1}

\labelformat{problem}{#1}

\newtheorem{theorem}{Theorem}
\newtheorem{problem}{Problem}

\newtheorem{lemma}{Lemma}
\newtheorem{remark}{Remark}

\newtheorem{definition}{Definition}

\newcommand{\sun}[1]{\textcolor{black}{#1}}

\newcommand{\feng}[1]{\textcolor{black}{#1}}

\newcommand{\yy}[1]{\textcolor{black}{#1}}

\graphicspath{{Images/}}
\usepackage{ulem}

\usepackage[marginal]{footmisc}

\begin{document}

\author{Tianfeng Feng}
\thanks{These authors contributed equally}
\affiliation{ QICI Quantum Information and Computation Initiative, Department of Computer Science, The University of Hong Kong, Pokfulam Road, Hong Kong}

\author{Tianqi Xiao }
\thanks{These authors contributed equally}
\affiliation{School of Physics, Sun Yat-sen University, Guangzhou 510275, China}
 
\author{Yu Wang}
\email{wangyu@bimsa.cn}
\affiliation{Yanqi Lake Beijing Institute of Mathematical Science and Applications, Beijing 101408, China}

\author{Shengshi Pang}
\affiliation{School of Physics, Sun Yat-sen University, Guangzhou 510275, China}

\author{Farhan Hanif}
\affiliation{Blackett Laboratory, Imperial College London, London SW7 2AZ, United Kingdom}
 
\author{Xiaoqi Zhou}
\email{zhouxq8@mail.sysu.edu.cn}
\affiliation{School of Physics, Sun Yat-sen University, Guangzhou 510275, China}

\author{Qi Zhao}
\email{zhaoqi@cs.hku.hk}
\affiliation{ QICI Quantum Information and Computation Initiative, Department of Computer Science, The University of Hong Kong, Pokfulam Road, Hong Kong}

\author{M. S. Kim}
\affiliation{Blackett Laboratory, Imperial College London, London SW7 2AZ, United Kingdom}

\author{Jinzhao Sun}
\email{jinzhao.sun.phys@gmail.com}
\affiliation{ Clarendon Laboratory, University of Oxford, Parks Road, Oxford OX1 3PU, United Kingdom}

\title{Two measurement bases are  asymptotically informationally complete for any pure state tomography}

\date{\today}

\begin{abstract}

One of the fundamental questions in quantum information theory is to find how many measurement bases are required to obtain the full information of a quantum state. While a minimum of four measurement bases is typically required to determine an arbitrary pure state, we prove that for any states generated by finite-depth Clifford + T circuits, just two measurement bases are sufficient. More generally, we prove that two measurement bases are informationally complete for determining algebraic pure states whose state-vector elements represented in the computational basis are algebraic numbers. Since any pure state can be asymptotically approximated by a sequence of algebraic states with arbitrarily high precision, our scheme is referred to as asymptotically informationally complete for pure state tomography. Furthermore, existing works mostly construct the measurements using entangled bases. So far, the best result requires $\mathcal{O}(n)$ local measurement bases for $n$-qubit pure-state tomography.  Here, we show that two measurement bases that involve polynomial elementary gates are sufficient for uniquely determining sparse algebraic states. Moreover, we prove that two local measurement bases, involving single-qubit local operations only, are informationally complete for certain algebraic states, such as GHZ-like and W-like states. Besides, our two-measurement-bases scheme remains valid for mixed states with certain types of noises. We numerically test the uniqueness of the reconstructed states under two (local) measurement bases with and without measurement and depolarising types of noise. Our scheme provides a theoretical guarantee for pure state tomography in the fault-tolerant quantum computing regime.

\end{abstract}

\maketitle

A fundamental question in quantum information theory is what is the minimum number of measurement bases required to determine an unknown pure quantum state. 
This is known as the Pauli problem which Pauli posed in 1933~\cite{pauli1933handbuch,pauli1958encyclopedia}: can an arbitrary wave function be uniquely determined (up to a global phase) by the probability distributions obtained from its position and momentum measurements? A generalisation of the Pauli problem is whether an unknown quantum state with dimension $d$ can be uniquely determined using two measurement bases ($d = 2^n$ for an $n$-qubit system). 
The answer to the generalised Pauli problem is negative. The number of measurement bases is shown to be larger than $3$ for $d \geq 9$~\cite{moroz1994problem}.
Subsequently, research progress has been made in understanding the requirements for achieving information completeness in high-dimensional pure states~\cite{mondragon2013determination,flammia2005minimal,goyeneche2015five,zambrano2020estimation,jaming2014uniqueness,sun2020minimal}.  Flammia \textit{et al.}~showed that a measurement with $2d - 1$ outcomes cannot be informationally complete and proved the failure of two complementary measurement bases for a unique determination of pure states~\cite{flammia2005minimal}.
From an information completeness perspective, a minimum of four measurement bases is {necessary}   for pure-state tomography~\cite{mondragon2013determination,jaming2014uniqueness}.

Although two measurement bases are not informationally complete for determining any pure-state, we prove that they are sufficient to determine any states generated by  Clifford + T circuits with finite depth {which can approximate any state up to arbitrary finite precision.}              
More generally, an unknown quantum state can be expressed as $|\psi \rangle =\sum_{k=0}^{d-1}{ \gamma_k}|k\rangle $, and we refer to the states with the coefficient $\gamma_k$ being an algebraic number for all $k$  as pure \textit{algebraic states}. We can readily find that any states generated from finite-depth Clifford + T circuits, and any constructed states with finite precision (e.g. states stored in classical memory in the task of pure state tomography), belong to the class of pure algebraic states.
In this work, we prove that two measurement bases are informationally complete for determining any pure algebraic state.
Our scheme can be regarded as a kind of quantum state tomography (QST) that involves the measurement of copies of an unknown quantum state using a set of designed measurements, which leverage the prior knowledge of the unknown quantum state, like matrix product state \cite{cramer2010efficient,qin2024quantum}, or permutation invariant state \cite{christandl2012reliable,gao2014permutationally}.
Here, we prove that the class of algebraic states is asymptotically equivalent to the whole pure state class.
Thus, two measurement bases are asymptotically informationally complete for pure-state tomography.

Following the minimal number of measurement bases from an information completeness perspective, \sun{ the next question is about the hardness in constructing and implementing the measurement bases}.  
For example, the construction of four measurement bases, saturating the lower bound required by information completeness, proposed in Ref.~\cite{jaming2014uniqueness} relies on Hermite functions and is challenging to realise.  Goyeneche \textit{et al.} and Zambrano \textit{et al.} proposed the five and three measurement bases strategy (with an invalid set), respectively. 
For local measurement bases,  so far a minimum of $mn+1$ ($m \geq 2$) separable bases are required for $n$-qubit pure-state tomography and it fails for sparse states \cite{pereira2022scalable}. In addition, Verdeli \textit{et al.} introduced a four-local-measurement scenario \cite{verdeil2023pure}. However, there is only numerical evidence without a theoretical validation of the solution \cite{pereira2022scalable,verdeil2023pure}.
Recently, Huang \textit{et al.} considered a relatively simpler task, i.e., state certification, and showed that almost all pure states could be efficiently certified through local measurements \cite{huang2024certifyingquantumstatessinglequbit}.
Other learning-based pure-state tomography protocols can be found in \cite{an2024unified,zhang2024variational}.
The second contribution of this work is that, along the line of general qudit state, we can readily find that for sparse algebraic states, two measurement bases that involve polynomial elementary gates (CNOT and single-qubit gates) are sufficient. 
Furthermore, we provide an explicit construction of the two measurement bases involving single-qubit local operations only, which we refer to as a {two-local-measurment-bases} scheme. Specifically,  the first basis is the computational basis, and the second basis is designed to be a rotated computational basis, which can be realised by applying a single-qubit Pauli rotation gate before measurements.
We prove that two local measurement bases are informationally complete for certain states such as GHZ- and W-like states.
Moreover, we show that for pure states {whose phases when expanded in computational basis are zero}, two local measurements are sufficient for strict information completeness, i.e., there does not exist any state (which could be non-algebraic) that can reproduce the same probability distribution under the two measurement bases.

Thus far, we have given an affirmative answer to the Pauli problem when concerning physically relevant limitations, showing that two measurement bases are asymptotically informationally complete for determining any pure states. 
We confront a few plausible questions and discuss clearly under which scenario our two measurement schemes apply unambiguously in the section of Discussion.
Moreover, we extend the discussion to mixed states with noise and prove that our {two-measurement-bases} scheme remains valid for mixed states with known types of noise, as, for instance, where the magnitudes of the noise coefficients and the noise type are known, the amplitude of the quantum state can be uniquely determined.
Finally, we validate our scheme {for both pure-state and mixed-state cases} through numerical simulations.
Numerical results manifest that the reconstructed quantum state generated from the measurement probability distributions is unique, as characterised by the high average fidelity over a large number of randomly generated states.
We test the effectiveness of the two-local-measurement-bases scheme for W-like states with up to 20 qubits.
Moreover, we incorporate measurement and device noise into the state and find that our scheme can be robust against noise.

\section{The Pauli Problem}

\vspace{10pt}
\textbf{Motivation.---}
Now we revisit and formulate the Pauli problem.
An arbitrary $d$-dimensional quantum pure state can be represented in the computational basis as 
\begin{equation}
    |\psi \rangle =\sum_{k=0}^{d-1}{\gamma_k }|k\rangle,
    \label{eq:psi_def_main}
\end{equation}
where the coefficient can be represented as $\gamma_k := r_k e^{i\varphi _k} $ with non-negative modulus $r_k\geqslant 0$, and the sign of the modulus is incorporated into the unknown phase $\varphi _k \in [0, 2\pi)$. 
Since the global phase factor has no physical observation effects, without loss of generality,  we set $\varphi _0=0$ hereafter. 
It is straightforward to see that a pure quantum state has  $2d-2$ free parameters. 
By performing measurements on the canonical basis, we can obtain $d$ measurement outcomes.
Thus, $2$ different measurements intuitively can provide $2 (d-1)$ equations (after removing the normalisation conditions), which aligns with the total number of unknown parameters of the quantum pure state. 
However, quantum interference introduces nonlinear terms into the equations derived from the two measured probability distributions, potentially resulting in multiple solutions and thereby hindering the unique determination of the pure state
\cite{flammia2005minimal}. \textbf{
The question is, what is the
largest set of state spaces within which two measurements are sufficient to determine the state?}

The quantum state generated from a finite-depth quantum circuit, indeed, is a subset of pure quantum states. 
For example, the quantum state generated by a finite number of Clifford and T gates can be expressed as: \( |\phi\rangle = \sum_{k=0}^{d-1} r_k e^{i \pi \varphi_k} \ket{k} \), where \( \varphi_k \in [0, 2\pi) \)  and \(r_k\) are all algebraic numbers. This structure ensures that all parameters of \( |\phi\rangle \) are algebraic numbers. 
This motivates us to consider a subset of pure states. To facilitate the discussion, let us put it formally by introducing a subset that is asymptotically equivalent to the full set.

\vspace{10pt}
\noindent \textbf{Asymptotic information completeness.---
}
We define the algebraic states as follows.
Suppose that the pure state can be represented in the computational basis in \autoref{eq:psi_def_main}.
A state is termed a pure \textit{algebraic state} if all coefficients $\gamma_k$ of the state are algebraic numbers.
The set of all pure states, denoted as $\mathbb{P}$,  consists of two subsets: $\mathbb{P}_A$ and $\mathbb{P}_{\overline{A}}$, satisfying $\mathbb{P}=\mathbb{P}_A \cup \mathbb{P}_{\overline{A}}$. Let $\mathbb{P}_A$  represent the set of algebraic states.
The complement set of $\mathbb{P}_A$, $\mathbb{P}_{\overline{A}}$, represents the set of pure states of which there exists at least one coefficient that is a transcendental number. Correspondingly, pure states in  $\mathbb{P}_{\overline{A}}$ are called \textit{transcendental states}. 

We prove that $\mathbb{P}_A$ can represent any states in $\mathbb{P}_{\overline{A}}$ with arbitrary precision. In other words, the subset of pure states $\mathbb{P}_A$ is asymptotically equivalent to $\mathbb{P}$.  
Our target problem is the following.

\begin{problem}
   What is the minimum number of orthogonal bases that are sufficient to uniquely determine a pure state $\ket{\psi} \in \mathbb{P}_A$ among all the \yy{other} pure states in $\mathbb{P}_A$.
   \label{problem_1_main}
\end{problem}

This is essentially an information-completeness problem. 
A set of measurement bases is informationally complete if any two distinct quantum states can be distinguished with confidence based on the outcomes of these measurements. 
To answer the question posed above, we frame it in the language of quantum information. 
For all $|\psi_1\rangle, |\psi_2\rangle \in \Lambda_A$,
these projections $\{|\phi_j^k\rangle\langle \phi_j^k|\}$ are called informationally complete with respect to a subset $\Lambda_A$ of all pure states $\mathbb{P}$ if the following condition is satisfied, 
$$
    \forall j,k,~~ |\langle \psi_1|\phi_j^k \rangle|^2= |\langle \psi_2|\phi_j^k \rangle|^2 \Rightarrow  |\psi_1\rangle=|\psi_2\rangle.
 $$ 
The above condition can be understood as follows. If any two states have the same probability distribution under \feng{any} measurements, they must be the same.

We can see the connection with QST, which involves the measurement of copies of an unknown quantum state using a set of designed measurements, followed by the reconstruction of the original state using the chosen algorithm \yy{\cite{flammia2005minimal,heinosaari2013quantum}}. In order to accurately reconstruct the original quantum state, the measurements used in the tomography should ideally be informationally complete with respect to a given set. This means that they should provide enough information to uniquely determine the state.  
Here, we focus on the set of algebraic pure states.
As we shall prove below, two measurement bases are asymptotically informationally complete for pure-state tomography. 
\sun{We would like to emphasize that this asymptotic informational completeness does not imply that increasing measurement precision to infinity would make two measurement bases sufficient for any pure state.} 
Instead, it means that two bases are sufficient for any state within the algebraic state space, which is asymptotically equivalent to the full pure-state space.

\section{Two  measurement bases scheme and its  asymptotic information completeness}
\label{sec:scheme_qudits}
 
\noindent\textbf{Scheme.---}
Now, we demonstrate that two measurement bases are sufficient to determine any pure algebraic states concerning the physical limitations.
The two measurement bases are illustrated in \autoref{fig:settings}.
First, we measure the algebraic states with two different measurement bases, and obtain two probability distributions $\mathbf{P}$ and $\mathbf{Q}$. 
Then the target pure quantum state can be reconstructed by post-processing the acquired distributions.
The first measurement basis is the computational basis, 
\begin{equation}\label{B0}
    \mathcal{B}_1 := \{ |i\rangle \},  i = 0,1,...,d-1.
\end{equation}
\sun{For the qubit case, that is the $z$-direction projective measurement $\mathcal{B}_1 := \{ |i\rangle \}^{\otimes n}, i = 0,1$. }
The second measurement basis is realised by applying a quantum Fourier transform $F$ for a qudit  and diagonal operation  $P$ 
before projective measurements on a computational basis. 
For an $n$-qubit state, this Fourier transform corresponds to local Hadamard gates $H^{\otimes n}$.
This can be  expressed as, 
\begin{equation}\label{B1}
     \mathcal{B}_2  :=  \{ S' | i \rangle\} ,  i = 0,1,...,d-1,
\end{equation}
where $S' = DF$ and $F$ is the standard quantum Fourier transform,  $F=\sum_{k,j=0}^{d-1} \frac{1}{\sqrt{d}} \omega^{jk}\ket{j}\bra{k}$ 
with $\omega =e^{\frac{2\pi i}{d}}$ and the diagonal operation is
 $D = \operatorname{Diag}\{ e^{i \theta_0}, ..., e^{i \theta_{d-1}}\}$ with angle $\theta_i$ as illustrated in \autoref{fig:settings}(a). {These two operations are experimentally feasible in quantum photonic systems, particularly on photonic chips \cite{zhu2022space,wang2020integrated}.}
 {For the qubit case, the operator is $S^\prime=DH^{\otimes n}$ illustrated in \autoref{fig:settings}(b) and could be simplified for sparse states which we will discuss later.
Since the global phase has no physical observation effects,  $\theta _0=0$ is set for convenience.

In order to satisfy the uniqueness property, we impose the following constraints on the  angles
\begin{equation}
\begin{aligned}
\forall i\ne j, i^\prime \ne j^\prime, |\theta_i -\theta_j|\ne |\theta_{i'} -\theta_{j'}|,
\label{eq:angles11}
\end{aligned}
\end{equation}
{where all $\theta_i$ $(i \leq d-1)$ are algebraic numbers.}
The choice of $P$ is not unique. Here we give a few examples that satisfy the above relationship.
{
A simple example satisfying the above relationship is
$ 
\theta_k = 2^{k-1} z_0
$
where $z_0$ is an arbitrary algebraic number for all $k \leq d-1$.} Another example is
$\theta_k=\sqrt{q_k}$ where $q_k$ is the $k$th prime number sorted from smallest to largest.
In this section, we simply give a measurement scheme for qudit. The uniqueness of the states under such a measurement scheme is presented in \autoref{thm:main}, illustrated in \autoref{fig:settings}(c).

\begin{figure}[t]
    \centering
    \includegraphics[width=0.5\textwidth]{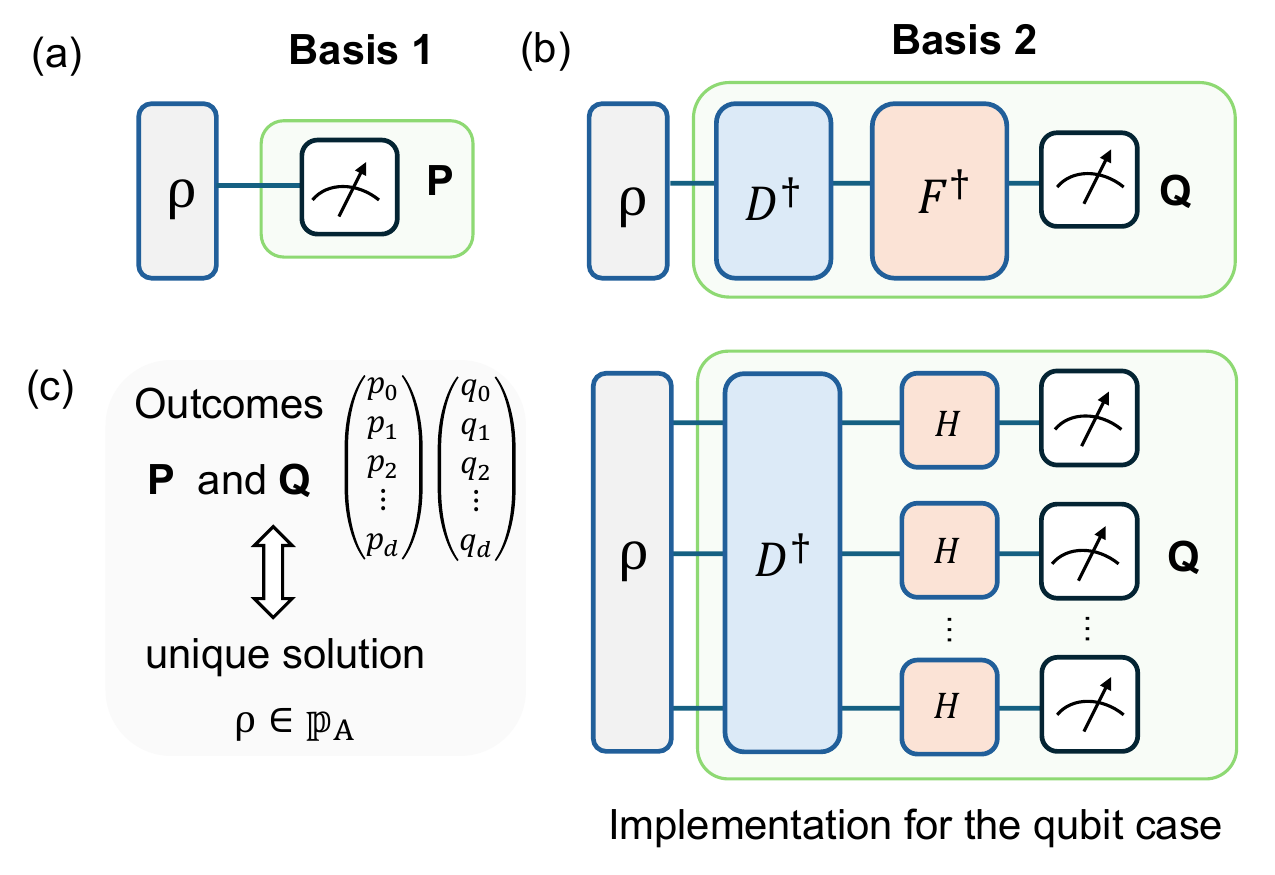}
    \caption{\textbf{Two-measurement-base scheme.} \textbf{a}. Scheme for a qudit state: $F$ is the standard quantum Fourier transform,  $F=\sum_{i,j=0}^{d-1} \frac{1}{\sqrt{d}} \omega^{ij}\ket{i}\bra{j}$, where $\omega =e^{\frac{2\pi i}{d}}$, and the diagonal operation is
 $D = \operatorname{Diag}\{ e^{i \theta_0}, ..., e^{i \theta_{d-1}}\}$ with angle $\theta_i$. \textbf{b}. Two measurement bases for multi-qubit systems. If $\rho$ is a sparse state, i.e. the number of non-zero components of $r_k$ is $\mathcal{O}(\text{poly}(n))$, $D$ can be reduced to be a diagonal operation with $\mathcal{O}(\text{poly}(n))$ phase shifts.
 \textbf{c}. Illustration of the concept of uniqueness: The pure quantum state $\in \mathbb{P}_A$ has a one-to-one correspondence to the measurement outcomes $\mathbf{P}$ and $\mathbf{Q}$.  
 }
    \label{fig:settings}
\end{figure}

\vspace{12pt}

\vspace{6pt}

\noindent
\textbf{Asymptotic information completeness.}
In this section, we establish the uniqueness of the probability distributions of an algebraic quantum state under our two measurement bases using the method of contradiction. 
\sun{
Briefly, any algebraic state generates unique probability distributions $\mathbf{P}$ and $\mathbf{Q}$.}

\begin{theorem}[Two   measurements are asymptotically informationally complete for any qudit and qubits states] 
The two measurement bases in \autoref{B0} and \autoref{B1} that satisfy \autoref{eq:angles11} are informationally complete with respect to all pure $d$-dimensional states $\in\mathbb{P}_A$. 
 
   \label{thm:main}
\end{theorem}

\sun{An equivalent expression is that the measurement outcomes on the two bases have a one-to-one correspondence with an algebraic state.
Our result does not violate the established information-theoretic conclusions as it focuses on information completeness with respect to a subset of the whole pure state class.}

Below, we briefly illustrate the proof idea of \autoref{thm:main} by contradiction and refer to  Supplementary Information for a complete proof.

\vspace{6pt}
\noindent \textbf{Sketch of the proof.---} Assume there exists another different algebraic quantum state that can give the same probability distributions of the target state under the two measurement bases (thus we call it a simulated state in this paper). As such, one can obtain a set of equations about the target state and the simulated state.
{We then apply the proved linear independence theorem to show that such a simulated state does not exist. }

Suppose that there exists such a $|\psi ^{\prime}\rangle \in \mathbb{P}_A$
can simulate the unknown algebraic state $|\psi \rangle $ under the computational basis, that is,  {$|{\langle j  \ket{\psi^{\prime}}}|^2=|{\langle j  \ket{\psi}}|^2$} for any $j$.
Therefore, without losing generality, we represent the states as $|\psi \rangle =\sum_{j=0}^{d-1}{r_je^{i\varphi _j}|j\rangle}$ and $|\psi ^{\prime}\rangle =\sum_{j=0}^{d-1}{r_je^{i\varphi _{j}^{\prime}}|j\rangle}$, with non-negative modulus $r_j$ in \autoref{eq:psi_def_main}.
By assumption, $|\psi ^{\prime}\rangle $ can also simulate the unknown algebraic state $|\psi \rangle $ under the second basis and have the same measurement probability distributions $\mathbf{Q}$ as $|\psi\rangle $. 
That is, the difference of the probability distributions for the second basis between $\ket{\psi}$ and $|\psi ^{\prime}\rangle $ is $\varDelta Q_j=Q_j-Q_{j}^{\prime} =0 $ for any outcome. 
Since $\mathbf{Q}$ corresponds to the probability distribution of the measurements under the basis $\{S^{\prime}\ket{j}\}$, $Q_j=|\langle j|S^{\prime\dagger} |\psi \rangle|^2 $ and $Q_j^\prime=|\langle j|S^{\prime\dagger} |\psi^\prime \rangle|^2 $. 
By grouping the terms, the identical distributions indicate that 

\begin{equation}
\begin{aligned}
\varDelta Q_j=\sum_{s<l \in V}C^{(j)}_{sl} \cos \left( \theta _s-\theta _l \right)+ S^{(j)}_{sl} \sin \left( \theta _s-\theta _l \right)=0,
\label{eq:paradox_main}
\end{aligned}
\end{equation}
where  $C^{(j)}_{sl} $ and $S^{(j)}_{sl}$
are the sum of the two trigonometric functions, 
\sun{which is a linear combination of cosine and sine of $ \varphi _s-\varphi _l$ and  $\varphi _{s}^{\prime}-\varphi _{l}^{\prime}$ (see Supplementary Information for more details).}

The proof of \autoref{thm:main} relies on the linear independence of sine and cosine functions: For a set of algebraic numbers $\mathcal{S} =  \{ \theta _k | |\theta_i| \neq |\theta_j|, \forall i, j \}$, $\{ \cos \theta _i \} $ and $\{\sin \theta _j\}$ are linearly independent over the set  $\mathcal{S} $ (see Lemma 1, Lindemann–Weierstrass theorem, in methods). 
Since our constraints of $\{\theta_j\}$ by \autoref{eq:angles11} indicate $|{\theta_s-\theta_l}|\ne |{\theta_m-\theta_n}|$ for $s,l\ne m,n$ and all $\theta_s-\theta_l$ are algebraic numbers. 
By \autoref{lemma:linear_indep} in Methods, there must be $ \varDelta Q_j\ne 0$
for all $j$, which contradicts the assumption that $\ket{\psi^\prime} \in \mathbb{P}_A$ can simulate $\ket{\psi}$. This completes the proof that  
two measurement bases are informationally complete for pure state tomography in $\mathbb{P}_A$.
As $\mathbb{P}_A$ is asymptotically equal to $\mathbb{P}$, our scheme is asymptotically informationally complete for pure-state tomography.

\vspace{6pt}



\vspace{6pt}
\noindent
{\textbf{Theoretical guarantee for multi-qubit pure sparse states.---}} 
Our above proof shows for any pure algebraic state, two measurement bases are informationally complete. 
However, as the state dimension increases, the complexity of implementing the second measurement basis also grows. For an $n$-qubit system with $d=2^n$, the implementation of a global phase gate may become exponentially complex in the general case \cite{nielsen_chuang_2010, Feng_2021}. 
Given the measurement distribution on the first basis, the implementation of the second basis can be significantly simplified. 
If the number of nonzero components of the quantum state in the computational basis scales polynomially with $n$, 
the operation $D^\dagger$ for the second basis can be simplified as follows:
\begin{equation}
 D^{\dagger} = \sum_{m=0}^{d-1}{e^{-i\theta _m(1-\delta_{r_m,0})} |m\rangle \langle m|}, \nonumber
\end{equation}
where $\delta_{r_m,0}$ is the delta function. 
Note that a single phase-shifting gate $\sum_{m\ne k} (\ket{m}\bra{m}+e^{-i\theta_k} \ket{k}\bra{k})$ can be implemented using a multi-qubit control phase gate (which can be realised by $\mathcal{O}(n)$ CNOT + single-qubit gates \cite{nielsen_chuang_2010}) and single-qubit NOT gates \cite{Feng_2021}. 
Thus, the gate complexity for a sparse $D^\dagger$ gate is $\mathcal{O}(\text{poly}(n))$.
The uniqueness of the sparse state $\in \mathbb{P_A}$ with simplified implementation also holds since the proof is the same as the general one.

\vspace{6pt}

\noindent \textbf{Discussions on the two-local-measurement-bases scheme.---}
{The above section mainly discussed the measurement scheme with entangled bases.} It is an open question whether a constant number of local bases is sufficient for information completeness \sun{with respect to $\mathbb{P}_A$}.
Here, we demonstrate that two local measurement bases are sufficient to determine a class of pure algebraic states, including GHZ-like states and W-like states.

Similar to the case of qudit, we first illustrate the processes of measurement and reconstruction in the multi-qubit setup.
First, we measure the algebraic states with two different local measurement bases, and obtain two probability distributions $\mathbf{P},\,\,\mathbf{Q}$. 
The first measurement basis is chosen as $\mathcal{B}_1$ in \autoref{B0} with $d=2^n$, which is the tensor product of projective measurements $\{|0\rangle\langle0|, |1\rangle\langle1|\}$ on each qubit. 
The second basis is constructed by
\begin{equation}\label{B1qubit}
   \mathcal{B}_2 := S'_q  \mathcal{B}_1,~\nonumber
\end{equation}
where $S'_q$ is a product of local rotations $S'_q = (R_nH)\otimes \cdots\otimes (R_1H)  $, $H$ is the Hadamard operation, $R_j$ is the phase operation 
\begin{equation}
R_j=\begin{bmatrix}
1 & 0 \\
0 & e^{i\alpha_j} \nonumber\\
\end{bmatrix}.
\end{equation}
 
The task is similar to the qudit case, which is to find the appropriate $\alpha_j$ such that they can uniquely determine a pure algebraic state.
However, the condition is not easily to be satisfied. This is because when restricting the locality of the measurement bases, the number of parameters is $n$ but the total number of linear equations is $2^n$.  Therefore, it is very likely that the parameters will become correlated, and thus, the linear independence among the equations will be lost. Therefore, there are multiple potential states that can have the same probability distribution and thus we cannot determine a unique state. 
The key is to ensure that their interference phase will not result in linear independence.
\sun{In this work, in order to ensure linear independence for the interference term in $\Delta Q_j$, we carefully design the conditions and propose a tentative solution.}

Now let us take a look at the global phase operation $\bigotimes_{j=1}^{n}R_j$. By expanding the tensor products of each diagonal operator $R_j$, we observe that this is a diagonal operation, which can be formally expressed as
$\bigotimes_{j=1}^n R_j=\sum_{m=0}^{2^n} e^{i \alpha_m' }\ket{m} \bra{m} $ with $\alpha _{m}^{\prime}=\sum_{p=1}^n{\alpha _pm_p}$ and
$|m\rangle =|m_n m_{n-1}\cdots m_1\rangle$.
Instead of restricting the absolute phase of the basis, an innovation here is that we propose to restrict its relative phase, i.e., $\alpha _{m}^{\prime}-\alpha _{k}^{\prime}$ for $m \neq k$.  Without loss generality,  any relative phase can be expressed 
$
  \sum_{m\in M}(-1)^{x_m} \alpha_m,
$
where $M$ is the subset of the set of internal numbers $N=\{1,2,3...,n\}$.
Given any two different strings $\Vec{x} \neq \Vec{x}'$ with $\Vec{x}=\{x_m\}$, the constraints on algebraic numbers $\{\alpha_m\}$ is
\begin{equation}
\begin{aligned}
\left|\sum_{m\in M}(-1)^{x_m} \alpha_m \right|\ne \left|\sum_{m\in M}(-1)^{x_m^\prime} \alpha_m \right|.
\end{aligned}
\label{eq_theta_qubits}
\end{equation}
This condition by \autoref{eq_theta_qubits} ensures that the linear independence theorem can be used in the analysis of information completeness of a set of pure states.

Although this condition looks more complicated than the qudit cases,  it is easy to find a set of measurement bases that satisfy this condition.
Similar to the qudit case, a simple example satisfying the above relationship is
$ 
\alpha _k=2^{k-1}z_0
$
 for any $k \leq n$, where $z_0$ is an arbitrary algebraic number.
Another example is $
\alpha _k=\sqrt{q_k}
$
where $q_k$ is the $k$th prime number sorted from smallest to largest.  

Compared to the qudit case, the measurement bases are local, making it easy to realise in experiments. 
We find that two local bases are sufficient to characterise special classes of sparse algebraic states: W-like algebraic states and GHZ-like algebraic states. Specifically, a general W-like  state and GHZ-like state are defined as follows respectively,
\begin{equation}
\begin{split}
      &  |\text{W-like}\rangle=\sum_{j=1}^n r_j e^{i\varphi_j} |0\rangle^{\otimes (j-1)} \otimes |1\rangle \otimes |0\rangle^{\otimes (n-j)},\\
&|\text{GHZ-like}\rangle =  \sum_{j=1}^n r_0 |0\rangle^{\otimes n} + r_1 e^{i\varphi} |1\rangle^{\otimes n} .~\nonumber
\end{split}
\end{equation}
We refer to Appendix \ref{local} for the proof.
It is worth noting that it is possible to identify other sparse algebraic states that can be characterised by two local bases and we leave the criteria in the Appendix.

\vspace{6pt}
\noindent \textbf{Discussions on strict information completeness.---}
The above section shows that two measurement bases are informationally complete with respect to $ \mathbb{P}_A$. Note that this result does not violate the minimum number of bases for pure-state tomography based on information-theoretic analysis.
In \autoref{thm:many_solution} we show that there exist multi-solutions belong to $\mathbb{P}_{\overline{A}}$ that can simulate the measurement outcomes $\mathbf{P}$ and $\mathbf{Q}$.
Finally, we conclude by proving a certain class of pure states can be uniquely determined by two local measurements.
Specifically, for a state $\ket{\psi} \in \Lambda $ defined in \autoref{eq:psi_def_main} whose relative phase is $0$ and amplitude $r_k$ is non-negative in computational basis and $ \Lambda$ being the set of such states, there exists no pure state $|\psi^\prime \rangle \in \mathbb{P}$, satisfying $|\psi^\prime \rangle  \ne |\psi\rangle  $, which can simulate the measurement outcomes of $\ket{\psi}$ in the bases $\{\ket{i}\}$ and $\{H^{\otimes n}|i\rangle\}$ (see Methods).
Therefore, for a certain class of pure states, including GHZ states, W states, and Dicke states, two local measurements are strictly informationally complete with respect to the whole set of pure states $\mathbb{P}$.

\begin{theorem}[For a certain class of pure states, two local measurements are {strictly} informationally complete ]
For a state $\ket{\psi} \in \Lambda $ defined in \autoref{eq:psi_def_main} whose relative phase is $0$ and amplitude $r_k$ is non-negative in computational basis and $ \Lambda$ being the set of such states, there exists no pure state $|\psi^\prime \rangle \in \mathbb{P}$, satisfying $|\psi^\prime \rangle  \ne |\psi\rangle  $, which can simulate the measurement outcomes of $\ket{\psi}$ in the two local bases $\{\ket{i}\}$ and $\{H^{\otimes n}|i\rangle\}$.
\label{thm:zero-phase}
\end{theorem}

If we only consider a set 
that is comprised of pure states in \autoref{eq:psi_def_main} with all zero phases, a single computational basis is sufficient to determine any state within this set. However,  one basis is insufficient if the set of the simulated states encompasses the entire pure state set. \autoref{thm:zero-phase} indicates that two local bases are sufficient to differentiate states with all zero phases within an entire set of pure states. For example, if Alice aims to verify the target state of a W state, but this quantum pure state is prepared for her by Bob, it would require measurements in two bases to determine whether the target state is indeed a W state. One basis involves a computational basis to confirm the absolute values of the state amplitudes, while the other basis is used to assess the phases and determine if they are all zero.
It is worth noting that although the relative phase of $\ket{\psi}$ is zero, we do not assume this is prior knowledge in the state tomography.
The introduction of the second basis can verify if the relative phase is zero.

\begin{table*}[ht]
\centering
\resizebox{1.0\textwidth}{!}
{
\begin{tabular}{c c c } 
 \hline
\hline
\textbf{Methods}     &  \textbf{Orthonormal Bases}  & \textbf{Requirements and Limitations}   \\
\hline
Flammia \textit{et al.} \cite{flammia2005minimal} &  $*$ & Requires at least $2d$ POVM elements or $3d-2$ rank-1 projectors; thus, any two orthonormal bases are insufficient. \\ 
   &  & Ineffective for some states in a measure zero set.  \\
    \hline
Finkelstein \textit{et al.} \cite{finkelstein2004pure} &   $*$ & Requires $2d$ rank-1 projectors, excluding a measure zero set.  \\
   &  & $3d-2$ rank-1 projectors are insufficient for all pure states.  \\
  \hline
Jaming \cite{jaming2014uniqueness} & 4   & Uses 4 eigenbases constructed mathematically for all pure states. \\ 
   & & Physically challenging to implement.  \\
    \hline
Carmeli \textit{et al.} \cite{carmeli2016stable} & 5 & Combines 4 eigenbases from \cite{jaming2014uniqueness} with the computational basis. \\
   & & No other mixed state matches the results, but physical implementation remains difficult.  \\
     \hline
Goyeneche \textit{et al.} \cite{goyeneche2015five} & 5    & Uses two entangled bases for multi-qubit systems.   \\
  &  & Ineffective for some states in a measure zero set.  \\
\hline
Zambrano \textit{et al.} \cite{zambrano2020estimation} & 3   & Guarantees at most $2^d-1$ candidate states, filtered by a likelihood function. \\ 
   & & Excludes a measure zero set.  \\
  \hline
Pereira \cite{pereira2022scalable}  &  $mn+1$  & Uses separable bases for multi-qubit systems, highly effective on NISQ quantum computers.  \\
   &  ($m\ge 2$) & Average estimation fidelity increases with $m$, but lacks theoretical guarantees.  \\
 \hline
Verdeil \textit{et al.} \cite{verdeil2023pure}  &  $2n+1$  & Utilizes Pauli measurements, suitable for NISQ quantum computers.  \\
  &  & Not applicable to qudit systems and lacks theoretical guarantees.  \\
\hline
\hline
\end{tabular}}
\caption{\textbf{Comparison of pure-state tomography methods.} This table summarizes various approaches for pure-state tomography, evaluating their applicability to pure states, implementation feasibility, and effectiveness in multi-qubit systems. All methods assume no prior knowledge of the target state. Note that all methods fail for some states in a measure zero set, except those addressed in \cite{jaming2014uniqueness,carmeli2016stable}.}
\label{table:EnergyComp}
\end{table*}

\vspace{6pt}
\noindent
\textbf{Understanding other works within this framework.---}
Previous results can be understood by this framework, where the state space is restricted to a certain set. The previous results show that at least four eigenbases, though they are hard to construct \yy{physically}, are informationally complete with respect to $\mathbb{P}$.
The original Pauli problem is whether a wavefunction is uniquely determined by its position and momentum probability distribution. Peres considered the discrete analogue problem of whether two eigenbases can uniquely determine all pure qudits except for a null set. The two eigenbasis are $\{|k\rangle\}_{k=0}^{d-1}$ and its Fourier basis $\{F|k\rangle\}_{k=0}^{d-1}$. That is to say, for some set  $C=\mathbb{P}-E$, two eigenbases are informationally complete with respect to $C$, where $E$ is the null set.
\sun{Efforts have been made along this line reducing the number of bases  \cite{goyeneche2015five,zambrano2020estimation}, and make it easy to construct by using local basis~\cite{verdeil2023pure}.
However, as shown in \cite{goyeneche2015five,flammia2005minimal,pereira2022scalable,zambrano2020estimation} are not effective for pure states in a null set.}
In this work, we consider the space with the algebraic states $\Lambda = \mathbb{P}$ and $\Lambda_A = \mathbb{P}_A$.
Two measurements are constructed, which are informationally complete with respect to all $d$-dimensional pure algebraic states. 
\sun{This aligns with the strategies for QST, in which prior knowledge of the target state is used to design the set of measurements (which need to be informationally complete).
}

\sun{
In existing works, the entangling bases in these works are non-orthonormal and may need to be adaptively changed according to the measurement outcomes, and they are not effective for determining certain states (in a set with measure zero). }

We summarise the comparison with other works in   Table~\ref{table:EnergyComp}.

\begin{figure*}[ht]
    \centering
    \includegraphics[width=\textwidth]{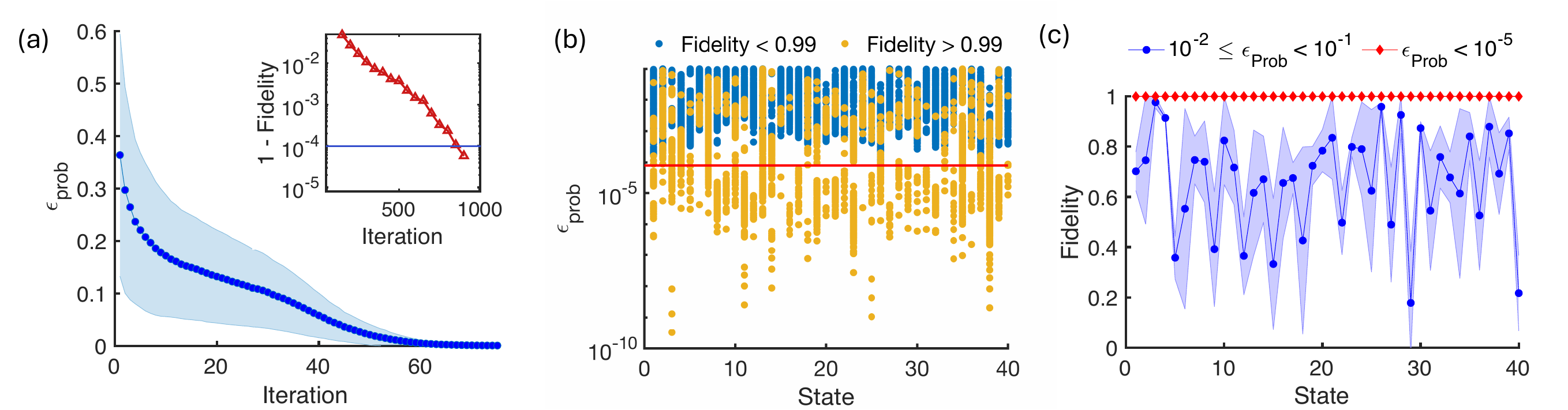}
    \caption{\textbf{Investigation on our local-measurement-bases scheme.} We study the error in probability distribution  $\varepsilon_{\textrm{Prob}}$ and the average fidelity and their relationship in the state reconstruction. 
    The algebraic numbers and transcendental numbers are simulated on a computer using low-precision numbers and high-precision numbers, respectively, as computers can only store data in finite decimal places. 
    The numbers of decimal places for the modulus $r_k$, the phases $\cos \varphi_k$ and  $\sin \varphi_k$ are denoted as $C_1$, $C_2$, and $C_3$, respectively, and are set to be $3$.
    The numbers of decimal places for the second measurement basis and the measured probability distributions are set to be $15$. The target state $\ket{\psi}$ and the simulated state $\ket{\psi'}$ are set at the same level of precision.
    \textbf{a}, {The error in probability distribution $\varepsilon_{\textrm{Prob}}$ in the reconstruction process for random states and 20-qubit W-like state.}
      We randomly generate 40 quantum states with dimension 8, and reconstruct each quantum state 2500 times using two local measurement bases scheme. \sun{To find the state, we use the Monte Carlo method to sample an initial guess and use the simulated annealing method to optimise $\varepsilon_{\textrm{Prob}}$. }The dark blue dots in the figure indicate the average values of $\varepsilon_{\textrm{Prob}}$, and the shaded area represent the standard deviation of the random states during the reconstruction.
     The figure inset shows the fidelity during the reconstruction process of the 20-qubit W-like state. 
    \textbf{b}, The horizontal coordinate is the 40 states we randomly generated and the vertical coordinate is the $\varepsilon_{\textrm{Prob}}$. The blue dots represent all reconstruction results with fidelity less than or equal to 0.99, while the orange dots represent all reconstruction results with fidelity greater than 0.99. The red line represents $\varepsilon _{Prob}=8\times 10^{-5}$.\textbf{c}, The horizontal coordinate is the 40 states we randomly generated, and the vertical coordinate is the fidelity between a target pure state and a reconstructed pure state. 
     The red and blue dots represent the average fidelity of all reconstruction results for a fixed target state with $\varepsilon_{\textrm{Prob}} < 10^{-5}$,  and $10^{-2}<\varepsilon_{\textrm{Prob}} < 10^{-1}$, respectively.
    The upper and lower bounds of the shaded area are the locations that deviate from the average values by one standard deviation. }
    \label{fig:M2_Numerous_results}
\end{figure*}

\begin{figure}[h!]
    \centering
    \includegraphics[width=0.9\linewidth]{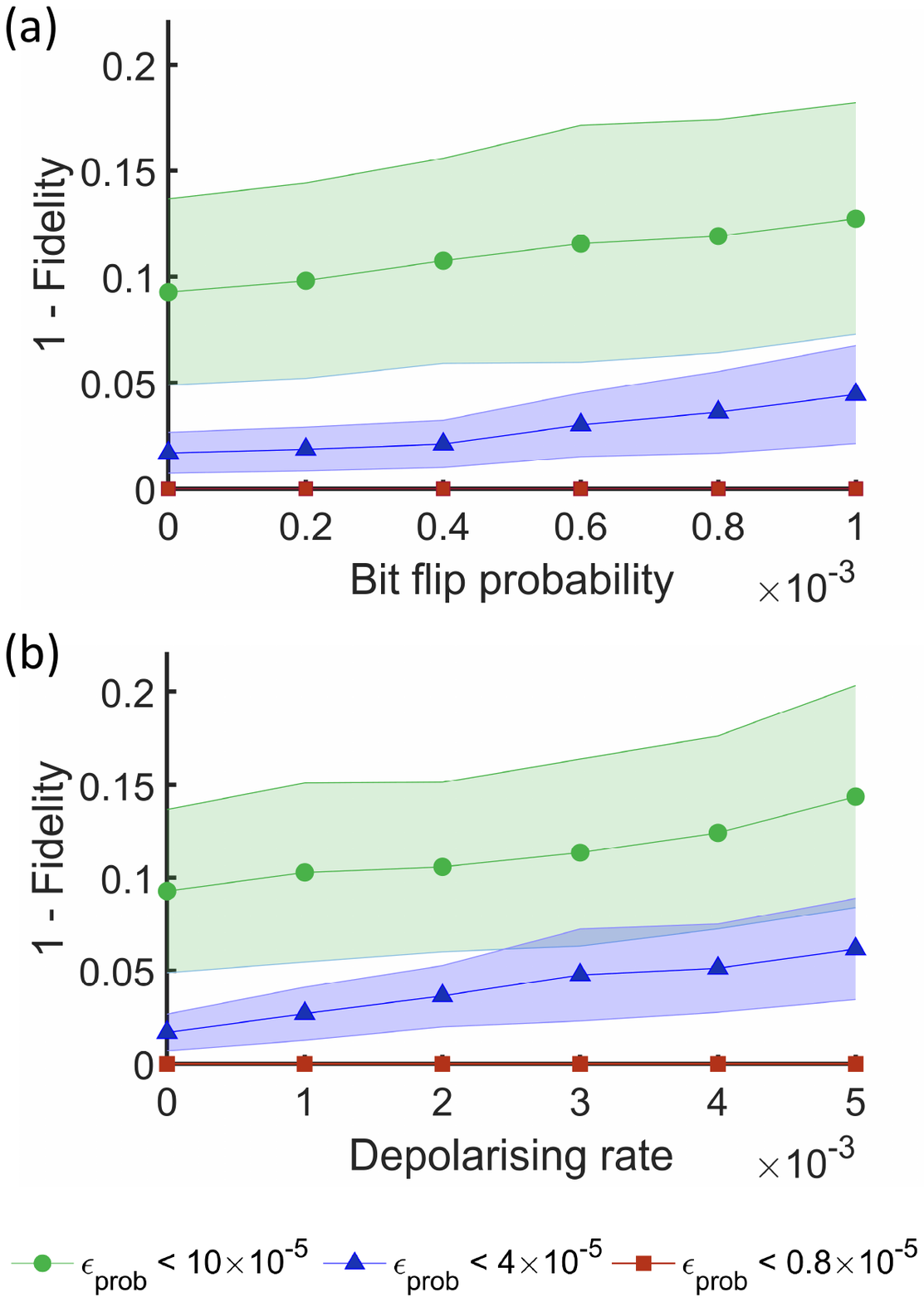}
    \caption{\textbf{Verification of our local measurement bases scheme with measurement and device statistical error.} 
    40 quantum states (of dimension 8) are randomly generated, and measurement results under two measurement bases are produced. Each state is reconstructed 2500 times.
    The setting for the decimal numbers is set the same as in \autoref{fig:M2_Numerous_results}.
   \textbf{a}, The bit-flip noise is added to the measurement results.\textbf{b}, The depolarising noise is added to the state.
    The red, blue, and green dots represent the average fidelity of all reconstruction results for a fixed target state with $\varepsilon_{\textrm{Prob}} < 0.8 
\times 10^{-5}$,  $\varepsilon_{\textrm{Prob}} < 4 \times 10^{-5}$, and $\varepsilon_{\textrm{Prob}} < 10^{-4}$, respectively.
     }
    \label{fig:Meas_Noise}
\end{figure}

\section{Numerical tests}


\sun{
In the above section, we have established rigorous proof of the asymptotic information completeness of two-measurement bases. Moreover, we have shown that \textbf{two local} measurement bases are sufficient for determining \feng{kinds of} sparse states. In this section, we focus on testing the effectiveness of reconstructing multi-qubit states obtained from purely local measurements. We numerically test the two-local-measurement-bases scheme from different aspects
}

Firstly, we consider the regimes where the target state is a pure state, and the ideal measurement probability distribution is obtained perfectly. 
The goal is to find quantum states that are as close as possible to the two sets of probability distributions $\mathbf{P}$ and $ \mathbf{Q}$.
We use the Monte Carlo algorithm and simulated annealing algorithm \cite{metropolis1953simulated,kirkpatrick1983optimization} to post-process the measured data obtained by our two-bases scheme as a way to obtain the amplitudes and phases of the reconstructed state.
We first generate a simulated state (a state whose matrix elements are randomly drawn from a certain distribution, and then the state is normalised) and compute its probability distributions $\mathbf{P}^{\prime}, \mathbf{Q}^{\prime}$ under our 2 measurement bases.
We define the distance between $\mathbf{P}^{\prime}, \mathbf{Q}^{\prime}$ and the actual measured probability distributions $\mathbf{P}$ and $ \mathbf{Q}$:
\begin{equation}
\begin{aligned}
\varepsilon_{\textrm{Prob}}=\left\| \mathbf{P}^{\prime}-\mathbf{P} \right\| _2+\left\| \mathbf{Q}^{\prime}-\mathbf{Q} \right\| _2,
\end{aligned}
\end{equation}
which is the error of probability distributions.
Small $\varepsilon_{\textrm{Prob}}$ simply means that the two states have similar performance when we observe from the measurement outcome, but it is not sufficient to verify the uniqueness of the state. Instead,  fidelity is a more faithful indicator of the uniqueness of the state. 
We will verify that the reconstructed quantum state is faithful as an indication of the uniqueness of the state. This faithfulness can be verified by the fidelity of the reconstructed state $
    \left| \langle \psi |\phi \rangle \right|^2
$.
Furthermore, we investigate the faithfulness of the reconstruction process by testing on an ensemble of states.
We investigate the relationship between $\varepsilon_{\textrm{Prob}}$ and the average fidelity. Our results indicate that when the $\varepsilon_{\textrm{Prob}}$ is lower than a certain value, the fidelity will be close to~$1$.

We start with the W-like states, which are typical sparse states that can be easily prepared in short depth,  
$ 
|\psi \rangle = \sum_{j=1}^n \gamma_j X_j  \ket{0}^{\otimes n}
$
with algebraic numbers $\gamma_j$ and $X_j$ being the Pauli X operator on the $j$th qubit.
\sun{It is worth noting that the local measurement bases are proven to be informationally complete for W-like states.}
The fidelity increase of the reconstructed states during the iteration process for the 20-qubit W-like state can reach more than $99.99\%$, and we leave the tomographic results in Supplementary Information.

\sun{To study the effectiveness of our local measurement scheme,} we choose to test with an eight-dimensional algebraic quantum state as our target state $|\psi \rangle$. 
The fidelity of the reconstructed state is $99.9998\%$ after 300 iterations, and can converge to $100\%$ by increasing the number of iterations.
Furthermore, we randomly generate 40 quantum states and show the reconstruction results in \autoref{fig:M2_Numerous_results}. A key finding is that when the $\varepsilon_{\textrm{Prob}}$ is small, the reconstructed quantum state has a fidelity approximating $100\%$. 
This indicates that our scheme can be verified if $\varepsilon_{\textrm{Prob}}$ is small enough.
We choose the iteration processes with $\varepsilon_{\textrm{Prob}}$ below $8\times 10^{-5}$ from all results and plot them in \autoref{fig:M2_Numerous_results}{(a)}.
In addition, we make two plots with data from the 100,000 reconstructed states in order to have a clear picture of the results, as shown in \autoref{fig:M2_Numerous_results}{(b)(c)}.
\autoref{fig:M2_Numerous_results}{(b)} shows that when the $\varepsilon_{\textrm{Prob}}$ is above $8\times 10^{-5}$, the fidelity can be both above and below $0.99$, but when the $\varepsilon_{\textrm{Prob}}$ is below $8\times 10^{-5}$, all the fidelity can be greater than $0.99$. \sun{In other words, this result excludes the existence of multiple solutions and shows the uniqueness of the state.}
\autoref{fig:M2_Numerous_results}{(c)} shows that the average fidelity is close to $100\%$ when the $\varepsilon_{\textrm{Prob}}$ is sufficiently small.
These simulation results illustrate that the reconstructed state can converge to the target state when the $\varepsilon_{\textrm{Prob}}$ of the reconstructed state is small enough, thus verifying the effectiveness of two bases. 
We further study the performance of our scheme with different levels of precision when dealing with the target states and the measurement bases. These results validate the necessity of introducing the algebraic space. Importantly, it shows the separation of the results for algebraic states and transcendental states (see Figure 4 in Methods).


To test the robustness of our measurement scheme, we add \feng{measurement bit-flip} noise with the noise rate $p$ in measurements, and construct the states from the noisy measurement outcomes.
Fig.~\ref{fig:Meas_Noise}{(a)} shows that the average fidelity of all reconstructed data decreases as the statistical noise increases in small increments.
If we observe the data with low $\varepsilon_{\textrm{Prob}}$, e.g., $\varepsilon_{\textrm{Prob}}$ below $0.8\times 10^{-5}$, the average fidelity of the reconstructed data maintains sufficiently high.
 These investigations provide a response to the limitations set up by our information-theoretic approach.

Finally, we numerically show that our scheme can be extended to mixed-state cases, where we add depolarising noise to the  target state $|\psi \rangle$ of the form
$
\rho =\left( 1-p \right) |\psi \rangle \langle \psi |+\frac{p}{d}I
 $
and measure $\rho$ under two measurement bases, where $p$ is the coefficient of the depolarising noise, $d$ is the dimension of the quantum states, and $I$ is the identity operator. 
We reconstruct pure quantum states from the probability distributions under two measurement bases using the same algorithms. 
The results are shown in \autoref{fig:Meas_Noise}{(b)}.
It can be seen that the average fidelity of all reconstructed data is decreasing as the depolarising noise rate increases in small increments.
However, when we analyse the data with $\varepsilon_{\textrm{Prob}}$ below $0.8\times 10^{-5}$, we can see that the fidelity of the reconstruction results are all close to $100\%$.
These results indicate that our scheme can be extended to mixed-state tomography.
\sun{In Methods, we study the performance when adding another measurement basis. Intuitively, adding more bases will make our scheme more robust because the states are more likely to be distinguished with more measurement results. We find that when the depolarising noise rate reaches 0.10, the fidelity can still be $95.61\%$, indicating that the addition of a measurement base helps to greatly enhance the noise immunity of our measurement scheme.
We leave more discussions to Methods.}

\section{Discussions}

This work {asks a fundamental} question of what the largest set of state spaces is for which two measurement bases are sufficient.
We find that the largest possible set is the algebraic state space, which includes, for example, states generated from the discretised fault-tolerant gate set (e.g. Clifford + T gate set) and any constructed states with finite precision which are concerned in the task of state tomography.
We rigorously demonstrate that two measurement bases are sufficient to determine any pure algebraic state (both qudit and qubit states).
\sun{Further, the two-local-measurement-bases scheme is {numerically investigated} for any pure algebraic qubit states.}
The two-measurement scheme is closely connected to the essence of quantum mechanics which requires complex amplitudes.
Even though the pure state itself can not be accurately achieved experimentally, we are interested in the heart of quantum mechanics and the pure state is mostly concerned with quantum mechanical problems in theory, as well as quantum states that can be generated from quantum circuits in the fault-tolerant regime.
We remark that the two measurements result only holds for determining $\ket{\psi}\in \mathbb{P}_A$, which has zero measure with respect to the full state space, but is asymptotically equivalent to the full set of pure states.
We show that a certain class of states is strictly informationally complete with respect to the full space under the two-local-measurement-bases scheme.

We extend the scheme to mixed states and provide the theoretical guarantee for the determination of mixed states with certain types of noise.
We validate our method for both pure states and mixed states through numerical simulations, where the reconstruction algorithms are simulated annealing algorithms with the Monte Carlo method.
We scale up numerical simulation by considering the reconstruction with a large problem size of up to 20 qubits.
The simulation results show that the quantum state can be uniquely constructed from the measurement outcomes in two local bases and the fidelity can reach more than $99.99\%$, and thus validates our information-theoretic result.
A critical question is whether the two-measurement scheme is still valid when considering a practical setup where measurement errors are present.  
Although the sufficiency of two bases in this practical situation is not proven, we investigate the robustness against measurement noise through numerics.
We observe that under measurement noise or depolarising noise, the reconstructed state can still achieve high fidelity, thus further validating the robustness of our scheme.

While this work focuses on the uniqueness of quantum state determination, our results can be applied to classical problems. 
The essence of the two-measurement scheme is to perform measurements on a computational base and a transformed Fourier base. Similar operations can be realised in the classical world: The Fourier transformation is widely used in classical problems such as image processing and analysis of communication systems.
For example, an image is placed in front of a positive lens and illuminated with parallel monochromatic coherent light, at which point the measurement on the back focal plane of the lens is the measurement of the image in the Fourier base~\cite{goodman2005introduction}.
Moreover, many problems at the classical level are treated discretely; for example, in image processing, the number of possible colours of a pixel is finite, which can be viewed as a result of high coarse-graining.
As such, this classic problem is a special case of \autoref{problem_1_main}, and we can directly apply our scheme to determine the target classical state.

\feng{For the foundation perspective,}
the fundamental formulae of quantum mechanics encompass two key facets: quantum states and their evolution. Initially, we approached these concepts by representing them as unit complex vectors and complex unitary operations, constraining our discussions within this framework. Our work suggests that through plausible simplifications by narrowing our focus to unit algebraic numbers and algebraic unitary operations, we may get better results. This refinement allows us to streamline our analysis while maintaining the essence of quantum mechanics. 
By removing the original constraints, we could open up new avenues of exploration; e.g., focusing on a subset of the pure state space may simplify other problems in quantum mechanics.

Our scheme can be considered a tomography scheme for the algebraic state using two measurement bases. The extension of our scheme to algebraic channel tomography is possible.  Besides,
an interesting problem is whether one can infer many properties of an unknown quantum state, such as quantum coherence \cite{coherenceRevModPhys.89.041003} and  entanglement~\cite{walter2013entanglement}, from just measurement outcomes of two bases without full state tomography.

\vspace{12pt}

\section*{Methods}

\subsection{Definitions}

We first show the asymptotic equivalence between $\mathbb{P}_A$ and $\mathbb{P}$ which are defined in the main text.

\begin{remark}[Asymptotic equivalence between $\mathbb{P}_A$ and $\mathbb{P}$]
\label{thm:equivalence}
    $\mathbb{P}_A$ can represent any states in $\mathbb{P}_{\overline{A}}$ with arbitrary precision. In other words, $\mathbb{P}_A$ is asymptotically equivalent to $\mathbb{P}$ if the precision is large enough.
\end{remark}

\begin{proof}

For any transcendental states $|\psi\rangle$ in the set $\mathbb{P}_{\overline{A}}$, we can construct a series of algebraic states $\{|\phi_n\rangle\}_{n=1}^{\infty}\in \mathbb{P}_A$ such that

\begin{equation}
   \lim_{n\to \infty} \| |\phi_n\rangle  - |\psi\rangle \| = 0.
\label{eq:sequence}
\end{equation}

The proof of \autoref{eq:sequence} is the following.
We express $|\psi\rangle\in\mathbb{P}_{\overline{A}}$ with its vector form $(\psi_0,\cdots,\psi_{d-1})$ under the computational basis. The state-vector norm $\| |\psi\rangle  - |\phi \rangle\|^2$ is defined as $\sum_{k=0}^{d-1} |\psi_k-\phi_k|^2$.  
For each transcendental number
 $\psi_k$, there always exists a sequence of algebraic numbers $\{b_{kj}\}_{j=1}^{\infty}$ to approximate it. 
 That is to say, for any given $\epsilon >0$, there always exists an integer $N_k$ such that, $|b_{kj}-\psi_k|^2< \epsilon /d$ when $j>N_k$. 
 Find the maximal number $N$ among $N_k$, $k=0,\cdots,d-1$. Thus we can construct a seriors of  unnormalised algebraic vector $|\tilde \psi_n\rangle=(b_{0n},\cdots,b_{d-1,n})$. The states $|\tilde \psi_n\rangle$ is approximate to $|\psi\rangle$ when $n\ge N$. 
 On the other hand, \sun{informally}, we know that $\sum_{k}|b_{kn}|^2 \to \sum_{k}|\psi_k|^2=1$ \sun{when $n\rightarrow \infty$}. Therefore, the normalised algebraic vector $| \psi_n\rangle=\frac{1}{\sum_{j}|b_{jn}|^2} |\tilde \psi_n\rangle$ is close to $|\tilde\psi_n\rangle$ \sun{when $n\rightarrow \infty$.}
Thus, the algebraic state $| \psi_n\rangle$ is close to vector $|\psi\rangle$ when $n$ is large enough.

\end{proof}

An operation $O$ is algebraic if \sun{its expression $ O_{jk} =  \braket{j|O|k}$ under basis $\mathcal{B}_1$ is algebraic for all indices $j,k=0,\cdots,d-1$. If both the state $|\phi\rangle$ and $O$ are algebraic, then $O|\phi\rangle$ is also algebraic}. Additionally, any finite sequence of Clifford + $T$ gates yields an algebraic unitary operation. 
In the regime of fault-tolerant quantum computing, it is natural to consider the states in $\mathbb{P}_A$ only.
Besides the tomography or certification of algebraic states, it may also be interesting to discuss the validation, decomposition, and error mitigation of algebraic operations.

\begin{remark}
Any state that is generated from the gate set of Clifford + T gates or circuits consisting of single-qubit rotating gates (the rotation angle should be $\beta \pi$, where $\beta$ is a rational number, and $\beta = 1/4$ for the T gate) + CNOT gates with finite depth belongs to the set~$\mathbb{P}_A$.
\end{remark}

Since we restrict the discussion to $\mathbb{P}_A$, a subset of $\mathbb{P}$, we introduce the following definition of information completeness.

\begin{definition}[Information completeness with respect to a given set~\cite{heinosaari2013quantum}] 
\label{def:information_complete_wrt}

Suppose the state space $\Lambda_A$ is a subset of state space $\Lambda$. 
Consider projective measurements onto several sets of eigenbases $\{|\phi_j^k\rangle\}$ for $j=0,1,...,d-1$. 

If the following condition is satisfied, 
 \begin{equation}
    \forall j,k,~~ |\langle \psi_1|\phi_j^k \rangle|^2= |\langle \psi_2|\phi_j^k \rangle|^2 \Rightarrow  |\psi_1\rangle=|\psi_2\rangle
 \end{equation}

1. for all $|\psi_1\rangle\in \Lambda$, and $|\psi_2\rangle\in \Lambda$,
these projections $\{|\phi_j^k\rangle\langle \phi_j^k|\}$ are called informationally complete with respect to~$\Lambda$.

2. for all $|\psi_1\rangle\in \Lambda_A$, and $|\psi_2\rangle\in \Lambda$,
these projections $\{|\phi_j^k\rangle\langle \phi_j^k|\}$ are called strictly informationally complete with respect to a subset $\Lambda_A$ of all pure states $\Lambda$.

3.
for all $|\psi_1\rangle,|\psi_2\rangle\in \Lambda_A$, these projections $\{|\phi_j^k\rangle\langle \phi_j^k|\}$ are called informationally complete with respect to a subset $\Lambda_A$ of all pure states $\Lambda$.

  \end{definition}

While all valid quantum pure states are unit vectors in a Hilbert space, the pure states that are typically manipulated in practical scenarios form a small subset of the entire Hilbert space. We define a subspace, referred to as the algebraic number space, which effectively covers the vast majority of quantum pure states encountered in practical applications and can approximate any quantum state to arbitrary precision. By selecting a computational basis within this space, along with an additional set of bases from its complement, the transcendental number space, these quantum pure states in the algebraic number space can be uniquely determined.

Intuitively, this problem \sun{and its solution} can be understood as bearing some resemblance to G\"{o}del's incompleteness theorem: in mathematics, certain ``truths'' always lie beyond the scope of a given system, much like how a language cannot fully encapsulate all the rules that govern itself. By leveraging the complementarity of algebraic and transcendental numbers, problems defined within the algebraic number space can be addressed in a broader mathematical framework. For instance, extending from the field of rational numbers to real numbers enables solutions to many previously unsolvable problems. Similarly, a set of bases in the transcendental space provides the capacity to encode all phases of quantum pure states defined in the algebraic number space. 
\vspace{10pt}
\subsection{Linear Independence Theorem}

Here we introduce two main lemmas for the proofs of our theorems.
\begin{lemma}[Lindemann–Weierstrass theorem \cite{lindemann1882zahl, weierstrass1885lindemann, nesterenko2021lindemann}]
    If $\alpha _1,\alpha _2,\cdots ,\alpha _m$ are distinct algebraic numbers, then the exponentials $e^{\alpha _1},e^{\alpha _2},\cdots ,e^{\alpha _m}$ are linearly independent over the field of all algebraic numbers.

\end{lemma}

\begin{lemma}[Linear Independence Theorem] 
Linear independence of sine and cosine functions with different angles over algebraic numbers: 
For a set of algebraic numbers $\theta _1,\theta _2,\cdots ,\theta _n$ that are not equal in absolute value to each other, $\cos \theta _1,\sin \theta _1,\cos \theta _2,\sin \theta _2,\cdots ,\cos \theta _n,\sin \theta _n$ are linear independent over the field of algebraic numbers.

\label{lemma:linear_indep}
\end{lemma}

\textbf{Lemma 2} can be proved by \textbf{Lemma 1}. Below, we provide the complete proof.

\textit{Proof}: For a set of algebraic numbers $\theta _1,\theta _2,\cdots ,\theta _n$ that are not equal in absolute value to each other, suppose $\cos \theta _1,\sin \theta _1,\cos \theta _2,\sin \theta _2,\cdots ,\cos \theta _n,\sin \theta _n$ are not linear independent over the field of algebraic numbers, which means that it must satisfy:

\begin{widetext}

\begin{equation}
\begin{aligned}
k_1\cos \theta _1+k_2\sin \theta _1+k_3\cos \theta _2+k_4\sin \theta _2+\cdots +k_{2m-1}\cos \theta _m+k_{2m}\sin \theta _m=0,
\end{aligned}
\end{equation}
while $k_1,k_2,\cdots ,k_{2m}$ are algebraic numbers that are not all zeros.
We can rewrite $\cos \theta _j,\sin \theta _j, j=1,2,\cdots ,m$ as:
\begin{equation}
\begin{aligned}
\begin{cases}	\cos \theta _j=\frac{1}{2}e^{-i\theta _j}+\frac{1}{2}e^{i\theta _j}\\	\sin \theta _j=\frac{i}{2}e^{-i\theta _j}-\frac{i}{2}e^{i\theta _j}\\\end{cases}.
\end{aligned}
\end{equation}
Then we get:
\begin{equation}
\begin{aligned}
\frac{k_1+ik_2}{2}e^{-i\theta _1}+\frac{k_1-ik_2}{2}e^{i\theta _1}+\frac{k_3+ik_4}{2}e^{-i\theta _2}+\frac{k_3-ik_4}{2}e^{i\theta _2}+\cdots \frac{k_{2m-1}+ik_{2m}}{2}e^{-i\theta _m}+\frac{k_{2m-1}-ik_{2m}}{2}e^{i\theta _m}=0.
\label{eq:SproofF}
\end{aligned}
\end{equation}
\end{widetext}
From the Lindemann–Weierstrass  theorem, it follows that if a set of algebraic numbers $\theta _1,\theta _2,\cdots ,\theta _n$ that are not equal in absolute value to each other, we can know that
\begin{equation}
\begin{aligned}
c_1e^{-i\theta _1}+c_2e^{i\theta _1}+\cdots +c_{2m-1}e^{-i\theta _m}+c_{2m}e^{i\theta _m}\ne 0,
\end{aligned}
\end{equation}
while algebraic numbers $c_1,c_2,\cdots c_{2m}$ are not all zeros. Then there is only one case satisfying the \autoref{eq:SproofF}, e.g. 
\begin{equation}
\begin{aligned}
c_{2j-1}=\frac{k_{2j-1}+ik_{2j}}{2}=0, c_{2j}=\frac{k_{2j-1}-ik_{2j}}{2}=0, 
\end{aligned}
\end{equation}
where $j=1,2,\cdots ,m$. Thus, one has
\begin{equation}
\begin{aligned}
k_l=0, l=1,2,\cdots ,2m.
\end{aligned}
\end{equation}
This result contradicts the hypothesis that $k_1,k_2,\cdots ,k_{2m}$ are not all zeros. As a result, for a set of algebraic numbers $\theta _1,\theta _2,\cdots ,\theta _n$ that are not equal in absolute value to each other, $\cos \theta _1,\sin \theta _1,\cos \theta _2,\sin \theta _2,\cdots ,\cos \theta _n,\sin \theta _n$ are linear independent over the field of algebraic numbers.\\

\subsection{Understanding Previous works within our framework}


Flammia \textit{et al.} proved that for any measure zero set $\mathbb{P}_1$, any two eigenbases are not informationally complete with respect to $C$.


Here we compare the results with the prior representative works.

The original concept of information completeness is for all quantum states. 
For a general  positive operator-valued measure (POVM) $\{E_k\}$, we call them informationally complete if the following condition holds, 
\begin{equation}
    \forall k,~~ \mbox{tr}(E_k \rho_1)= \mbox{tr}(E_k \rho_2) \Rightarrow  \rho_1=\rho_2,
 \end{equation}
for all $\rho_1,\rho_2 \in \mathbb{A}$, where $\mathbb{A}$ corresponds to the set of all density matrices. This needs the measurement operators $\{E_k\}$ to span the whole Hermitian matrix space. 
Thus an informationally complete POVM should contain at least $d^2$ operators. It is proved that $d+1$ mutually unbiased bases (MUBs) are minimal and optimal. The symmetric informationally complete POVM (SIC-POVM) contains $d^2$ measurement outcomes \cite{scott2006tight}. But the existence of $d+1$ MUBs and SIC-POVM for arbitrary dimension $d$ is still unknown up to now. The existence problems are selected as the first two open questions of quantum information theory by Horodecki \textit{et al.} \cite{horodecki2022five}.  

In order to decrease the number of measurement outcomes, prior knowledge of the unknown state is taken into consideration. That is to say, the target unknown states are some subset of the whole space $\mathbb{A}$. The prior information could be matrix product states \cite{cramer2010efficient,qin2024quantum}, permutation invariant
states \cite{christandl2012reliable,gao2014permutationally}, and subsets from a fiducial state \cite{li2016fisher} and so on. 

Pure states correspond to the rank-1 density matrix. 
When the set for prior information of pure states is restricted $\mathbb{P}-E$, where $E$ is a set of measure zero, Flammia \textit{et al.} 
proved that any POVM  with $2d-1$ outcomes is not enough. Two special kinds of POVM are constructed: One is with $2d$ elements and the other is with $3d-2$ rank-1 elements \cite{flammia2005minimal}. Later, it is proved that POVM with $2d$ rank-1 elements is also enough \cite{finkelstein2004pure}. It is proved that five projective measurements onto orthonormal bases are also enough \cite{goyeneche2015five}. Adding an auxiliary qubit, the number of orthonormal bases can decrease to 2 or three, while the number of total measurement outcomes is the same or $4d$ \cite{wang2022pure}. 
Three orthonormal bases are designed for this task, where totally $2^d-1$ different pure states match the probability distributions. \cite{zambrano2020estimation}. For dimension $d=4$, it is proved that infinite candidates can match the probability distribution of Peres's two unbiased bases \cite{sun2020minimal}. For $n$-qubit systems, $mn+1$ separable bases \cite{pereira2022scalable} and $2n+1$ Pauli observables are considered \cite{verdeil2023pure}.  

 When the set of prior information of pure states is $\mathbb{P}$, it is proved that the POVM should contain at least $4d-3-c(d)\alpha(d)$ elements, where $c(d)\in [1, 2]$ and $\alpha(d)$ is the number of ones appearing in the binary expansion
of $d-1$ \cite{heinosaari2013quantum}. While the four orthonormal bases are theoretically enough~\cite{jaming2014uniqueness}. 
There is also a hot topic about directly reconstructing pure quantum states. With an auxiliary pointer and sequential weak measurements, two unbiased bases are enough~\cite{lundeen2011direct}. The ancilla can be removed from $\sigma$-quench measurement~\cite{zhang2019delta}. 

For certain schemes, the reconstruction method entails utilizing convex optimization to derive a density matrix that best matches the measurement data. 
Then theoretically, it should avoid another density matrix that matches the data of pure states. 
This needs the POVM $\{E_k\}$ to be strictly informationally complete with pure states, i.e.,
\begin{equation}
    \forall k,~~ \mbox{tr}(E_k |\phi\rangle\langle \phi|)= \mbox{tr}(E_k \rho) \Rightarrow  |\phi\rangle\langle \phi|=\rho,
 \end{equation}
for all $|\phi\rangle \in \mathbb{P}$ and $\rho \in \mathbb{A}$.
To ensure strict information completeness with pure states, five eigenbases are sufficient~\cite{carmeli2016stable}. At least $5d-7$ operators should be considered \cite{chen2013uniqueness}.  For 2-qubit pure states, a total of 11 Pauli measurements are required, while for 3-qubit pure states, 31 Pauli measurements are needed \cite{ma2016pure}. Pure states correspond to rank-1 density matrices. The definition of strict information completeness with pure states is generalised to strictly rank-$r$ complete \cite{baldwin2016strictly}, by replacing the set $\mathbb{P}$ with $\mathbb{A}_r$, where $\mathbb{A}_r$ corresponds to the set of all rank-$r$ density matrices. 
The compressed sensing method shows that randomly chosen $O(rn^2 2^n)$ Pauli observables can reconstruct unknown rank-$r$ density matrix with high probability \cite{gross2010quantum}. 

In our work, we directly construct the amplitudes and phases of unknown pure states using measurement data, resulting in the reconstruction of pure states rather than density matrices. During the calculation of phases ($\theta$),  $e^{i\theta}$ can be a transcendental number in theory. To include this case, we introduce \autoref{def:information_complete_wrt} of strict information completeness concerning a subset $\mathbb{P}_A$ of all pure states $\mathbb{P}$.

1. For all $|\psi_1\rangle,|\psi_2\rangle\in \mathbb{P}-E$, where $E$ is a measure zero set, any 2 sets of eigenbases are not informationally complete~\cite{flammia2005minimal}. 


2. For all $|\psi_1\rangle,|\psi_2\rangle\in \mathbb{P}$, the minimal number of eigenbases is 3 for $d=2$. When $d=3$ or $d\ge 5$, the minimal number is 4. The problem is still open for $d=3$, where 4 sets of eigenbases are enough by not clear whether 3 sets of eigenbases are also enough \cite{carmeli2015many}. 

3. For all $|\psi_1\rangle,|\psi_2\rangle\in \mathbb{P}-E$, where $E$ is a measure zero set, 5 sets of eigenbases are constructed \cite{goyeneche2015five}, $mn+1$ sets of eigenbases \cite{pereira2022scalable} and $2n+1$ Pauli bases are enough.

Another relevant topic is pure state verification. Rather than reconstructing unknown states belonging to a subset, the aim is to ascertain whether the states produced by a device are, on average, sufficiently close to the given target state $|\phi\rangle$. Various types of pure states are explored, including bipartite pure states, stabiliser states, hypergraph states, weighted graph states, Dicke states, answer vectors of linear equations, and more \cite{hayashi2006study,pallister2018optimal,zhu2019optimal,wang2019optimal,hayashi2015verifiable,markham2020simple,takeuchi2018verification,somma2021complexity,yu2019optimal}.

State learning: In the quantum state certification task, the goal is to determine whether a given set of $n$ copies of an unknown $d$-dimensional quantum mixed state $\rho$ matches a known mixed state $\sigma$, or if it differs from $\sigma$ by at least $\epsilon$ \cite{buadescu2019quantum}. 
\sun{Direct-fidelity estimation, classical shadow tomography, randomized benchmarking.} \feng{add more statements or delete}

\subsection{Proof of \autoref{thm:zero-phase}}

 \begin{proof}
Without loss generality, we suppose $\ket{\psi}=\sum_{i=0}^{2^n-1} r_i |i\rangle \in \Lambda $ where $r_i $ is real number and $r\ge 0$. 
Obviously, if $\ket{\psi^{\prime}}\in \mathbb{P}$ can simulate the measurement results of $\ket{\psi}$ under computational basis $\{\ket{i}\}$, $\ket{\psi^{\prime}}$ should have the following form:
 $\ket{\psi^\prime}=\sum_{i=0}^{2^n-1} r_i e^{i\phi_i} |i\rangle \in \mathbb{P} $  where $\{\phi_i\} \in [0,2\pi)$ are  undefined. For ease of presentation, we denote the $j$th measurement outcome of the $\{H^{\otimes n}|i\rangle\}$ basis for $\ket{\psi}$ and $\ket{\psi^{\prime}}$ as $T_j$ and $T_j^\prime$ respectively. 
 If $\ket{\psi^\prime}$ can produce the same measurement outcomes $T$ of $\ket{\psi}$, one has $T_j=T_j^{\prime} $,
 that is

\begin{equation}
\begin{aligned}
    \Delta T_j &= T_j-T_j^\prime \\
    &=\frac{1}{2^{n-1}}\sum_{s < l} (-1)^{h(j,s)+h(j,l)}r_s r_l [ 1-\text{cos} (\phi^{\prime}_s-\phi_l^{\prime})]=0.
\end{aligned}
\end{equation}
For $j=0$, $ 
    \Delta T_0= 2^{-(n-1)} \sum_{s < l}r_s r_l [ 1-\text{cos} (\phi^{\prime}_s-\phi_l^{\prime})].
$
Since $r_s r_l\ge 0$ and  $1-\text{cos} (\phi^{\prime}_s-\phi_l^{\prime})\ge 0$, $\Delta T_0\ge 0$. $T_0=0$ if and only if all $\phi_i^\prime=0$, i.e., $\ket{\psi}=\ket{\psi^{\prime}}$. Therefore, there is no $\ket{\psi^{\prime}}\in \mathbb{P}$ and $\ket{\psi^{\prime}} \ne \ket{\psi}$ that can simulate $\ket{\psi}$.

\end{proof}

\subsection{Discussions on strict information completeness}


Suppose the to-be-determined target state is $\ket{\psi} \in \mathbb{P}_A$. 
In the above section, we have shown that two measurement bases are informationally complete with respect to $ \mathbb{P}_A$, but they are not strictly informationally complete.
The following theorem shows that there exist multi-solutions belong to $\mathbb{P}_{\overline{A}}$ that can simulate the measurement outcomes $\mathbf{P}$ and $\mathbf{Q}$.


\begin{theorem}[The existence of multi-solutions with the same measurement outcomes, i.e., not strictly informationally complete with respect to $\mathbb{P}_A$]
    1.      There exists $\ket{\psi}\in \mathbb{P}_A$ with the measurement outcome $P$ and $Q$, and there exist many states $\ket{\psi'} \in \mathbb{P}_{\overline{A}}$ such that the measurement outcomes of $\ket{\psi'} $  are  $P$ and $Q$. 
    
    2. There exist infinitely many states $\ket{\psi}$ such that $\ket{\psi'} \in \mathbb{P}_{\overline{A}}$ has the same measurement outcomes as $P$ and $Q$, 
    and $|\braket{\psi' | \psi}| < 1 - \beta$ with a nonvanishing $\beta$.  
    \label{thm:many_solution}
\end{theorem}

Here we give a simple example to illustrate \autoref{thm:many_solution}. Consider the $n$-qubit GHZ-like state as the unknown target state
    \begin{equation}
    |\psi \rangle =a_0|0\rangle ^{\otimes n}+a_{2^n-1}e^{i\varphi _{2^n-1}}|1\rangle ^{\otimes n}.
    \end{equation}
There is one solution in $\mathbb{P}_{\overline{A}}$ that may meet $\mathbf{P}$ and $\mathbf{Q}$ if two measurement bases are given. The solution is $|\psi ^{\prime} \rangle =a_0|0\rangle ^{\otimes n}+a_{2^n-1}e^{i(-\varphi _{2^n-1}+2\alpha _{2^n-1}^{\prime})}|1\rangle ^{\otimes n}$, where $\alpha _{2^n-1}^{\prime}$ is the parameter corresponding to $\mathbf{Q}$ basis.
Obviously, this solution may be far away from $\ket{\psi}$. This is rooted in the fact that the quantum states in $\mathbb{P}_{\overline{A}}$,  \autoref{lemma:linear_indep}  no longer apply.
The proof is shown in the Appendix.

\autoref{thm:many_solution} indicates that there are many distinct states that can simulate the target state. Universal quantum computation only requires the implementation of circuits such as Clifford + T circuits, or circuits consisting of single-qubit Pauli rotation gates (the rotation angle should be $\beta \pi$, where $\beta$ is a rational number and T gate is $\beta = 1/4$) + CNOT gates. These circuits still make the evolving quantum state in $\mathbb{P}_A$, within which our scheme is valid,  \sun{although the set of these states has a zero measure compared to the full set of pure states}.  
Finally, we conclude by showing a certain class of pure states can be uniquely determined by two local measurements, that is, two local measurements are strictly informationally complete with respect to $\mathbb{P}$ for a certain class of pure states.

\subsection{Numerical setup, algorithm, and simulation}

\subsubsection{Generating algebraic states}

In the numerical simulation, the algebraic numbers and transcendental numbers are simulated on a computer using low-precision numbers and high-precision numbers, respectively, because computers can only store data in finite decimal places.
We set that the target states are represented using numbers with $M_1$ decimal places, while the simulated states are represented using numbers with $M_2$ decimal places.
The $e^{i\alpha}$ in the single-qubit rotation gates of the second basis are transcendental numbers with $X$ decimal places in target states, and $X^{\prime}$ decimal places in simulated states.
The numbers in probability distributions of target states $\bf P,\,\,Q$ are numbers with $Y$ decimal places, and probability distributions of simulated states $\bf P^{\prime},\,\,Q^{\prime}$ with $Y^{\prime}$ decimal places.
We require the measurements to be represented in high precision, and thus $X$, $X^{\prime}$, $Y$ and $Y^{\prime}$ are set to be greater than $M_1$ and $M_2$.
It is worth noting that if we do a truncation of ${M}$ decimal places on the target state or the simulated state directly, it will result in the final reconstructed quantum state not being unique.
In this case, the truncation error of amplitude normalisation is $O\left( 2^n10^{-M} \right)$, which grows exponentially as the number of qubits $n$ increases.
However, if we consider amplitude normalisation and phase normalisation, this error can be reduced, where a unique reconstructed state can be obtained.

\begin{figure*}[ht!]
    \centering
\includegraphics[width=\linewidth]{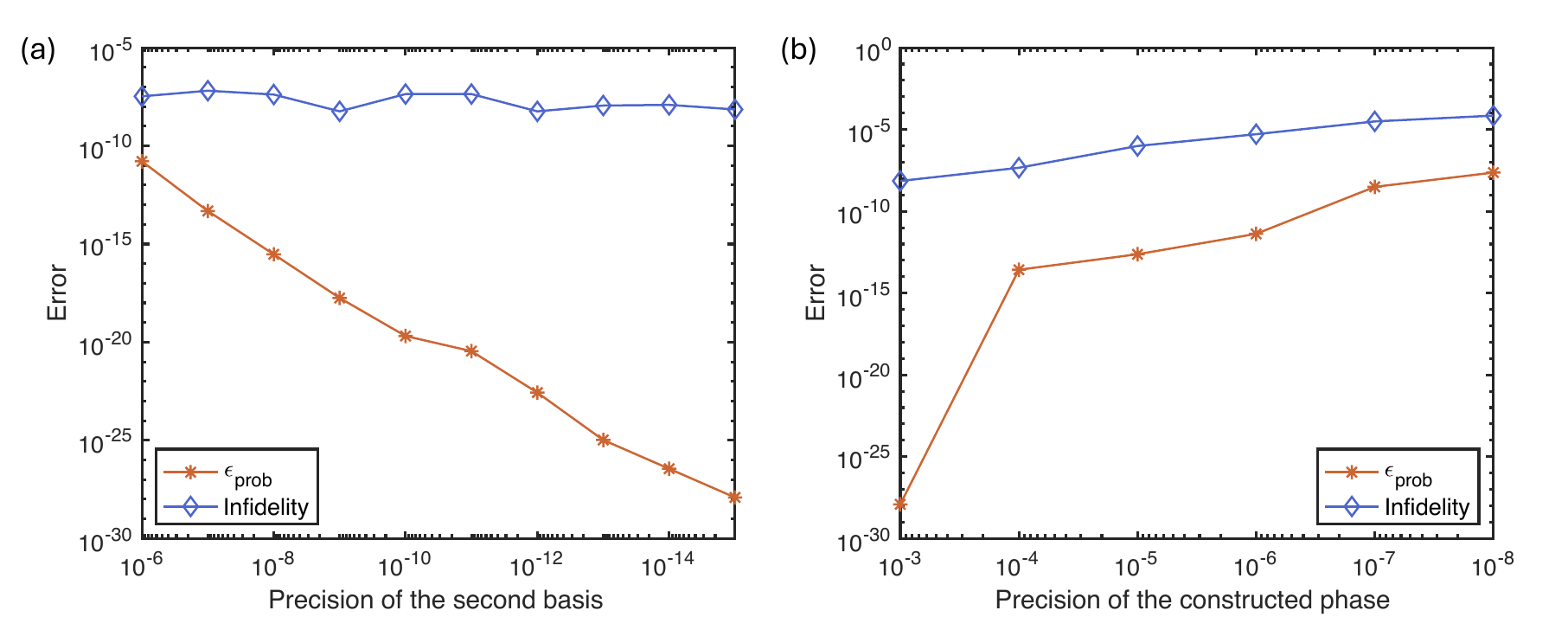}
    \caption{The number of decimal places of the second measurement base $X^{\prime}$ and the simulated phases $C_2^{\prime}$ during the reconstruction process are varied respectively, in order to test their effects on the $\varepsilon_{\textrm{Prob}}$ and average fidelity of the reconstructed 3-qubit quantum state by using two local measurement bases scheme. The numbers of decimal places for the amplitudes are $C_1=C_{1}^{\prime}=3$, for the phases are $C_2=3$ and $C_{2}^{\prime}$ changing from $3$ to $8$. The numbers of decimal places for the second measurement base are $X=15$ and $X^{\prime}$ changing from $6$ to $15$, and for the measured probability distributions are $Y=Y^{\prime}=15$. Forty 3-qubit quantum states are randomly generated and each quantum state is reconstructed 2500 times. (a)(b) We choose the reconstruction result with the smallest $\varepsilon_{\textrm{Prob}}$ for each state and average the $\varepsilon_{\textrm{Prob}}$ and fidelity over the 40 states.\textbf{a}, The horizontal coordinate is the number of decimal places of the second measurement base $X^{\prime}$, the blue dots reflect the values of $\varepsilon_{\textrm{Prob}}$, while the red dots reflect the values of the absolute values of 1 minus the average fidelity.\textbf{b}, The horizontal coordinate is the number of decimal places of the simulated phases $C_2^{\prime}$, the blue dots reflect the values of $\varepsilon_{\textrm{Prob}}$, while the red dots reflect the values of the absolute values of 1 minus the average fidelity.}
    \label{fig:M3_Decimal_places}
\end{figure*}

\begin{figure}[ht]
    \centering
    \includegraphics[width=\linewidth]{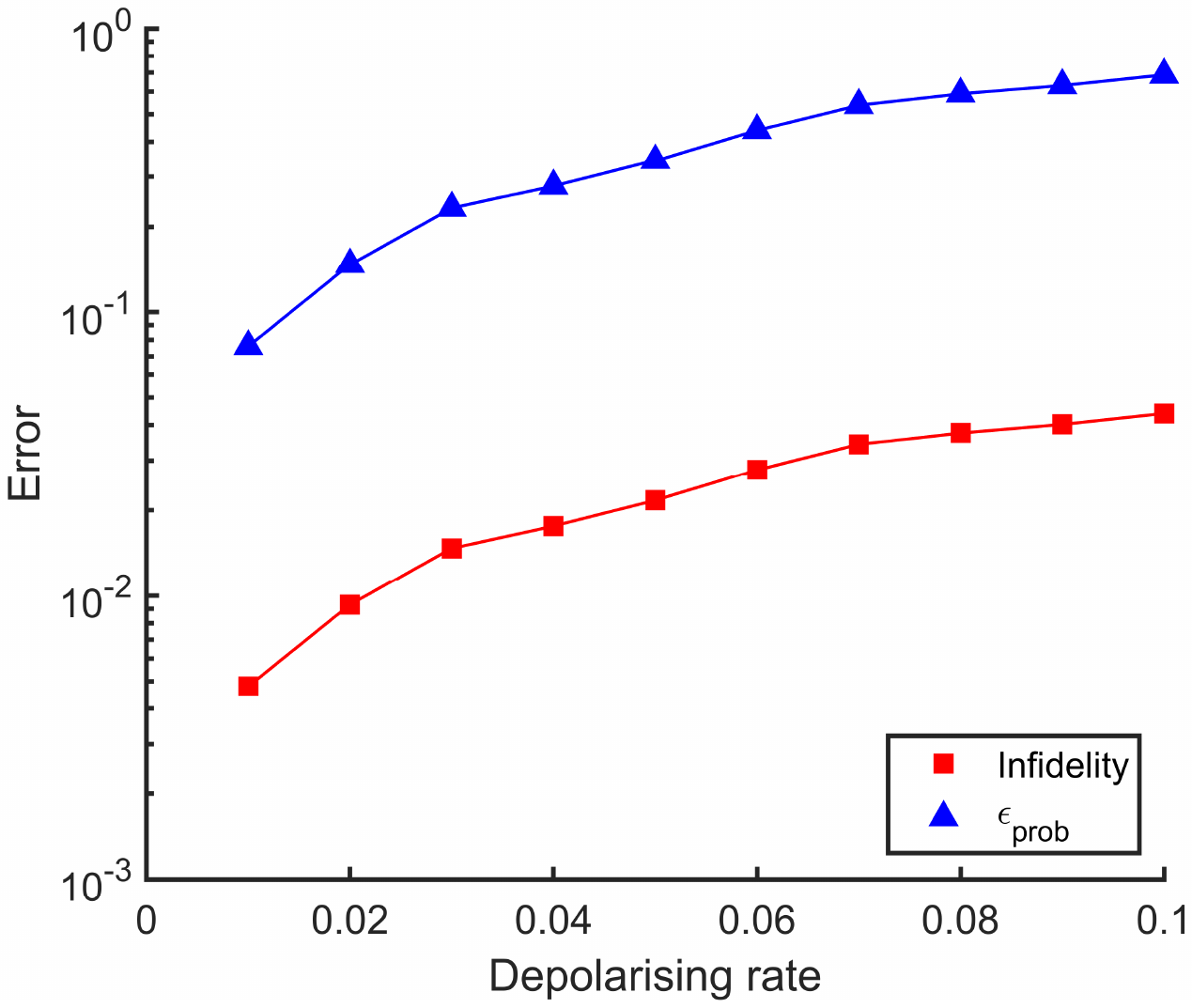}
    \caption{\textbf{Fidelity improvement when adding the third basis.} 
    A 5-qubit pure state is randomly generated and depolarising noise is added to produce its probability distributions under three measurement bases. Then we use the measurement results to reconstruct the pure state. The horizontal coordinate is the coefficient of the depolarising noise added to the randomly generated pure state, the blue dots reflect the values of $\varepsilon_{\textrm{Prob}}$, while the red dots reflect the values of the absolute values of 1 minus the fidelity.}
    \label{fig:third_basis}
\end{figure}

The target state can be expressed as
\begin{equation}
\begin{aligned}
|\psi \rangle =\sum_{k=0}^{2^n-1}{r_ke^{i\varphi _k}|k\rangle}=\sum_{k=0}^{2^n-1}{\left( r_k\cos \varphi _k+ir_k\sin \varphi _k \right) |k\rangle},
\end{aligned}
\end{equation}
and the simulated state can be similarly expressed as
\begin{equation}
\begin{aligned}
|\phi \rangle =\sum_{k=0}^{2^n-1}{r_{k}^{\prime}e^{i\varphi _{k}^{\prime}}|k\rangle}=\sum_{k=0}^{2^n-1}{\left( r_{k}^{\prime}\cos \varphi _{k}^{\prime}+ir_{k}^{\prime}\sin \varphi _{k}^{\prime} \right) |k\rangle}.
\end{aligned}
\end{equation}
In our setup,  we do truncation on $r_k,~ \cos \varphi _k, ~\sin \varphi _k\,\,\left( r_{k}^{\prime}, ~\cos \varphi _{k}^{\prime}, ~\sin \varphi _{k}^{\prime} \right)$, with the decimal places of the truncation set to $C_1, C_2, C_3\,\,\left( C_{1}^{\prime}, C_{2}^{\prime}, C_{3}^{\prime} \right)$. In this work, $C_2=C_3 = C_{2}^{\prime}=C_{3}^{\prime}$ is set. 
It is obvious that $M_1=C_1+C_2, M_2=C_{1}^{\prime}+C_{2}^{\prime}$.
In addition, the quantum state needs to satisfy amplitude normalisation and phase normalisation in \autoref{eq:52} and \autoref{eq:53}.

\subsubsection{
Algorithm for pure state reconstruction}

The goal is to find quantum states that are as close as possible to the two sets of probability distributions $\mathbf{P}$ and $ \mathbf{Q}$.
To do so, we use the Monte Carlo algorithm and simulated annealing algorithm \cite{metropolis1953simulated,kirkpatrick1983optimization} to post-process the measured data in our two-bases scheme as a way to obtain the amplitudes and phases of the reconstructed state.
We first generate a simulated state (a state whose matrix elements are randomly drawn from a certain distribution, and then the state is normalised) and compute its probability distributions $\mathbf{P}^{\prime}, \mathbf{Q}^{\prime}$ under our 2 measurement bases.
Recall that we calculate the error of probability distributions by $\varepsilon_{\textrm{Prob}}$.

If the $\varepsilon_{\textrm{Prob}}$ of this simulated state is less than the $\varepsilon_{\textrm{Prob}}$ of the last randomly generated simulated state, the stored result is updated to the current simulated state.
Otherwise, if the $\varepsilon_{\textrm{Prob}}$ of this simulated state is greater than or equal to the $\varepsilon_{\textrm{Prob}}$ of the last randomly generated simulated state, the stored result is updated to the current simulated state with a probability $p$, and the stored result is not updated with a probability $1-p$.
Repeat the above steps enough times.
With a sufficiently large number of repetitions, this update scheme will eventually converge to the quantum state corresponding to the minimum value of the $\varepsilon_{\textrm{Prob}}$ \cite{liu1996metropolized,liu2001monte}.

The classical post-processing for constructing the target states is efficient for sparse states.
We consider sparse states with $k$ nonzero amplitudes.
The measurement results of the computational basis allow us to obtain the amplitude coefficients and to know the position of the nonzero component of the quantum state.
In this case, the number of unknown phases is $k$ instead of $2^n  $.
Based on this analysis, we can reconstruct the $k$ unknown phases by post-processing the measurement results of the second measurement basis.

We consider the effect of the number of decimal places of the second measurement base $X^{\prime}$ on the $\varepsilon_{\textrm{Prob}}$ and the average fidelity during the reconstruction process (see Fig.~\ref{fig:M3_Decimal_places}{(a)}). 
We also consider the effect of the number of decimal places of the simulated phases $C_2^{\prime}$ on the $\varepsilon_{\textrm{Prob}}$ and average fidelity during the reconstruction process (see Fig.~\ref{fig:M3_Decimal_places}{(b)}).
As we can obtain the quantum state amplitude directly from the first basis measurement, i.e., $\mathbf{r}_{\mathbf{d}}=\sqrt{\mathbf{P}_{\mathbf{d}}}$, the number of decimal places of the simulated amplitudes $C_1^{\prime}$ has almost no effect on the $\varepsilon_{\textrm{Prob}}$ and average fidelity.
Also, classical computers store floating-point numbers with $15$ decimal places, so   $Y=Y^{\prime}=15$ is set.
As can be seen in both graphs, the error in probability distribution $\varepsilon_{\textrm{Prob}}$ fluctuates in a small range as the number of decimal places changes, but the average fidelity is stable around $99.999\%$.

\subsubsection{Adding more measurement bases.}
Moreover, we study the performance when adding another measurement basis.
Intuitively, adding more bases will make our scheme more robust because the states are more likely to be distinguished with more measurement results.
We find that adding a measurement base, such as
$ 
\left\{ |k\rangle \right\} =\left\{ H^{\prime}|i\rangle \right\} 
$, $H^{\prime}=H^{\otimes n}, i=0,1,\cdots ,2^n-1$
where $H$ is the Hadamard gate, the robustness of our scheme against noise can be greatly enhanced.
We randomly generate a 5-qubit pure state $|\psi \rangle$ and add depolarising noise. Then we obtain the probability distributions under three measurement bases, and post-process the noisy data.
We extend the definition by
$$\varepsilon_{\textrm{Prob}}=\sum_{i = 1}^3 \left\| \mathbf{P}_i^{\prime}-\mathbf{P}_i \right\| _2,$$ where $\mathbf{P}_{\mathbf{1}},\mathbf{P}_{\mathbf{2}},\mathbf{P}_{\mathbf{3}}$ represent the probability distributions under three measurement bases of the target states, and $\mathbf{P}_{\mathbf{1}}^{\prime},\mathbf{P}_{\mathbf{2}}^{\prime}$, and $\mathbf{P}_{\mathbf{3}}^{\prime}$ represent the probability distributions under three measurement bases of the simulated states.
Another measure of the quality of the reconstruction process is the  infidelity defined as
$ \left| 1-Fidelity \right|=\left| 1-\left| \langle \phi |\psi \rangle \right|^2 \right|
$.
Fig.~\ref{fig:third_basis} shows $\varepsilon_{\textrm{Prob}}$ and $\left| 1-Fidelity \right|$ versus the depolarising noise rate.
When the depolarising noise rate reaches 0.10, the fidelity can still be $95.61\%$, indicating that the addition of a measurement base helps to greatly enhance the noise immunity of our measurement scheme.

\subsection{Comments}
This work is mainly concerned with information completeness and is not concerned with measurement problems.
Intuitively, the uniqueness is easy to hold with respect to a relatively small set of the full pure state space.
\sun{In a more practical setup,  when the level of control in the precision of the single-qubit rotation for measurement settings is higher than that of the gate operations in the circuit, the reconstructed state is unique.} 
Moreover, the set of states generated by discretised gate sets (e.g. Clifford + T gates) is a subset of $\mathbb{P}_A$, and it is not a continuous set compared to the $\mathbb{P}_A$. Therefore, compared to the general case, it is easier to find the appropriate rotation angles such that the state can be uniquely constructed. 

The proof of uniqueness is based on the linear independence of the equations obtained from measurements, which is purely information-theoretic. 
The sufficiency of two bases hinges on the fact that (1) ideal measurement probability distribution can be obtained perfectly, and (2) the measurement bases can be set to have transcendental angles.
With these perfect assumptions, we will not encounter measurement issues in which statistical errors have to be considered.
A critical question is whether the two-local-measurement-bases scheme is still valid in a practical setup with measurement errors.
Although we cannot prove the sufficiency of two bases in this practical situation, we will investigate this point through numerics. 
We will take the statistical errors into account and show that our method can still faithfully reconstruct the state uniquely. These investigations provide a response to the limitations set up by our information-theoretic approach.

 \normalem


\begin{acknowledgments}
  The authors thank Ying Li and Qiming Ding for helpful discussions and suggestions. 
Q. Z. and T. F. acknowledge funding from National Natural Science Foundation of China (NSFC) via Project No. 12347104 and No. 12305030, Guangdong Natural Science Fund via Project 2023A1515012185, Hong Kong Research Grant Council (RGC) via No. 27300823, N\_HKU718/23, and R6010-23, Guangdong Provincial Quantum Science Strategic Initiative No. GDZX2303007, HKU Seed Fund for Basic Research for New Staff via Project 2201100596. M.S.K, J.S. and F.H. thank the UK EPSRC through EP/Y004752/1, EP/W032643/1, and the KIST Open Research Program. M.S.K also thanks the NRF grant (RS-2024-00413957). 
J.S. thanks support from the Innovate UK (Project No.10075020).
This research also received support through Schmidt Sciences, LLC.
X. Z. thanks the National Natural Science Foundation of China (Grant No. 61974168) and the National Key Research and Development Program (Grant No. 2017YFA0305200).  Y. W. thanks the National Natural Science Foundation of China (Grants No. 62001260 and No. 42330707) and the Beijing Natural Science Foundation (Grant No. Z220002).
\end{acknowledgments}

\widetext

\newpage
\appendix

\section*{Supplementary Information for "Two measurement bases are  asymptotically informationally complete for any pure state tomography" }



\section{Proof of the uniqueness for the qudit case in \autoref{thm:main}}

\subsection{Derivation of probability distributions for a qudit state}

A general qudit state can be written as
\begin{equation}
\begin{aligned}
|\psi \rangle 
=\sum_{k=0}^{d-1}{r_ke^{i\varphi _k}}|k\rangle,
\end{aligned}
\end{equation}
where $\sum_{k=0}^{d-1}{r_{k}^{2}=1}$ and  $r_k\geqslant 0$. 
Clearly, one can determine all $r_k$ based on the results of the probability distribution of the computational basis (the first measurement base in our scheme).
The unitary of the second measurement base of our scheme is given as
\begin{equation}
\begin{aligned}
S^{\prime\dagger}=F^{\dagger}D^{\dagger},
\end{aligned}
\end{equation}
where $F=\sum_{m,n=0}^{d-1}\frac{1}{\sqrt{d}}e^{\frac{2\pi i mn}{d}}\ket{m}\bra{n}$ is the $d$-dimensional quantum Fourier transform and 
$D^{\dagger}=\sum_{m=0}^{d-1}{e^{-i\theta _m}|m\rangle \langle m|}$ which is a diagonal phase gate.  Since $D^\dagger$ is a diagonal matrix, it alters only the phase of the quantum state, leaving the amplitude unchanged. Consequently, any qudit state will retain its form, apart from the phase shift induced by $D$. To ease the discussion, we begin by examining the action of   $F^\dagger$ on a general qudit state.\\

A quantum state $|\psi \rangle
=\sum_{j=0}^{d-1}{r_ke^{i\varphi _k}}|k\rangle$ after applying $F^{\dagger}$ is given by
\begin{equation}
\begin{aligned}
|\psi ^F\rangle =F^{\dagger}|\psi \rangle =\sum_{j=0}^{d-1}{\psi _{j}^{F}|j\rangle},
\end{aligned}
\end{equation}
where
$\psi _{j}^{F}=\frac{1}{\sqrt{d}}\sum_{k=0}^{d-1}{r_ke^{i\varphi _k}e^{-\frac{2\pi ijk}{d}}}=\frac{1}{\sqrt{d}}\sum_{k=0}^{d-1}{r_ke^{i\left( \varphi _k-\frac{2\pi jk}{d} \right)}}$.
The $j$-th amplitude of the state now becomes
\begin{equation}
\begin{aligned}
Q_{j}(F^\dagger\ket{\psi})=\left| \langle j|\psi ^F\rangle \right|^2=\psi _{j}^{F}\psi _{j}^{F\dagger}=\frac{1}{d}+\frac{2}{d}\sum_{s<l}{r_sr_l\cos \left[ \varphi _s-\varphi _l-\frac{2\pi j\left( s-l \right)}{d} \right]}.
\label{eq:ProbF}
\end{aligned}
\end{equation}
Since Eq. \ref{eq:ProbF} holds for a general state,  the probability distribution when measuring in our second base is  
\begin{equation}
\begin{aligned}
Q_j\left( F^\dagger D^\dagger |\psi \rangle \right) =Q_{j}\left( F^\dagger \ket{\phi}\right)  =\frac{1}{d}+\frac{2}{d}\sum_{s<l}{r_sr_l\cos \left[ \left( \varphi _s-\varphi _l \right) -\left( \theta _s-\theta _l \right) -\frac{2\pi j\left( s-l \right)}{d} \right]},
\label{eq:Prob2F}
\end{aligned}
\end{equation}
 where 
$|\phi \rangle =P^{\dagger}|\psi \rangle=\sum_{k=0}^{d-1}{e^{-i\theta _k}|k\rangle \langle k|\psi \rangle}=\sum_{j=0}^{d-1}{r_je^{i\left( \varphi _j-\theta _j \right)}|j\rangle}$.
This is because the measurement result under the second basis is equivalent to shifting the phase of the quantum state from $\varphi _j$ to $\varphi _j-\theta _j$ in the \autoref{eq:ProbF}.

\subsection{Proof of uniqueness of qudit states $\in \mathbb{P}_A$}

In our scheme, the vectors $\mathbf{P}$ and $\mathbf{Q}$ represent the measurement distributions of the two bases for an unknown quantum state $|\psi \rangle=\sum_{j=0}^{d-1}{r_je^{i\varphi _j}}|j\rangle$. Specifically, $P_j = r_j^2$ denoting the $j$th component of $\mathbf{P}$ and $Q_j=Q_j\left( F^\dagger D^\dagger |\psi \rangle \right)$ (see Eq. \ref{eq:Prob2F}) denoting the $j$th component of $\mathbf{Q}$. 
In the following, we prove the uniqueness of our two measurement bases for any algebraic pure state (i.e. $r_j$, $\cos(\phi_j)$ and  $\sin(\phi_j)$ are algebraic numbers) using a proof by contradiction. 

Specifically, suppose there exists one or more algebraic states $|\psi' \rangle\in \mathbb{P}_A$ that is different from the target algebraic state $|\psi \rangle\in \mathbb{P}_A$, yet has the same probability distributions $\mathbf{P}$ and $\mathbf{Q}$ in our two measurement bases as the target state $|\psi \rangle$.
The set consisting of all $|\psi' \rangle$ is called the null space, indicating that the tomography scheme will fail within this null space. 

Let us analyze the conditions that $|\psi' \rangle \in \mathbb{P}_A$ should satisfy.
First $|\psi ^{\prime}\rangle $ can simulate the unknown algebraic state $|\psi \rangle $ under the computational base. 
Without losing generality, we set $|\psi \rangle =\sum_{j=0}^{d-1}{r_je^{i\varphi _j}|j\rangle}$ and $|\psi ^{\prime}\rangle =\sum_{j=0}^{d-1}{r_je^{i\varphi _{j}^{\prime}}|j\rangle}$, where $r_j\geqslant 0$, $\varphi ,\varphi ^{\prime}\in \left[ \left. 0,2\pi \right) \right. $.
Given that the global phase has no physical observation effect, we first choose $g$ from the set $V=\{k\left|r_k\ne 0 \right. \}$, where $k$ is an integer that varies from $0$ to $d-1$ and simultaneously satisfies $r_k \ne 0$. Here, $g$ represents the smallest ordinal number of the non-zero amplitudes.
We then set its corresponding phase $\varphi_g = 0$ and $\varphi_g' = 0$, which means that we use it as the reference standard to define the remaining phases. It is important to note that if the amplitude of a component is zero, the value of its phase is meaningless.
Therefore a more accurate description of the unknown quantum state should be 
\begin{equation}
\begin{aligned}
|\psi \rangle =\sum_{j\in V}{r_je^{i\varphi _j}|j\rangle}=\sum_{j\in V,\, j\ne g}{r_je^{i\varphi _j}|j\rangle}+r_g|g\rangle.
\label{eq:fai1F}
\end{aligned}
\end{equation}
The corresponding quantum state $|\psi ^{\prime}\rangle $ that simulates the measurement result of unknown quantum state $|\psi \rangle $ under the computational base is
\begin{equation}
\begin{aligned}
|\psi ^{\prime}\rangle =\sum_{j\in V}{r_je^{i\varphi^\prime_j}|j\rangle}=\sum_{j\in V,\, j\ne g}{r_je^{i\varphi _{j}^{\prime}}|j\rangle}+r_g|g\rangle. 
\label{eq:fai2F}
\end{aligned}
\end{equation}

We have assumed that $|\psi ^{\prime}\rangle $ is different from $|\psi \rangle $ at the beginning of our proof, so at this point, there is at least one different phase with nonzero amplitude between $\ket{\psi}$ and $\ket{\psi^\prime}$.

Let us analyse whether $|\psi ^{\prime}\rangle $ can simulate the unknown algebraic state $|\psi \rangle $ under the second basis. If it is true, we have $\varDelta Q_j=Q_j-Q_{j}^{\prime}=0 $, where $Q_j^\prime
$ donates the measurement outcome of the simulated state $\ket{\psi
^\prime}$. 
Let us calculate the difference of the probability, e.g.
\begin{equation}
\begin{aligned}
\varDelta Q_j=Q_j-Q_{j}^{\prime}=\frac{2}{d}\sum_{s<l \in V}{r_sr_l\left\{ \cos \left[ \left( \varphi _s-\varphi _l \right) -\left( \theta _s-\theta _l \right) -\frac{2\pi j\left( s-l \right)}{d} \right] -\cos \left[ \left( \varphi _{s}^{\prime}-\varphi _{l}^{\prime} \right) -\left( \theta _s-\theta _l \right) -\frac{2\pi j\left( s-l \right)}{d} \right] \right\}}.
\label{eq:paradoxF}
\end{aligned}
\end{equation}
Expanding $\varDelta Q_j$ yields:
\begin{equation}
\begin{aligned}
\varDelta Q_j&=\frac{2}{d}\sum_{s<l\in V}{r_sr_l}\left\{ \cos \left[ \varphi _s-\varphi _l-\frac{2\pi j\left( s-l \right)}{d} \right] \cos \left( \theta _s-\theta _l \right) +\sin \left[ \varphi _s-\varphi _l-\frac{2\pi j\left( s-l \right)}{d} \right] \sin \left( \theta _s-\theta _l \right) \right. 
\\
&\quad\quad\quad\quad\quad\quad\quad\quad\quad\quad \left. -\cos \left[ \varphi _{s}^{\prime}-\varphi _{l}^{\prime}-\frac{2\pi j\left( s-l \right)}{d} \right] \cos \left( \theta _s-\theta _l \right) -\sin \left[ \varphi _{s}^{\prime}-\varphi _{l}^{\prime}-\frac{2\pi j\left( s-l \right)}{d} \right] \sin \left( \theta _s-\theta _l \right) \right\} 
\\
&=\frac{2}{d}\sum_{s<l \in V}{r_sr_l}\left\{ \left\{ \cos \left[ \varphi _s-\varphi _l-\frac{2\pi j\left( s-l \right)}{d} \right] -\cos \left[ \varphi _{s}^{\prime}-\varphi _{l}^{\prime}-\frac{2\pi j\left( s-l \right)}{d} \right] \right\} \cos \left( \theta _s-\theta _l \right) \right. 
\\
&\quad\quad\quad\quad\quad\quad\quad\quad\quad\quad +\left. \left\{ \sin \left[ \varphi _s-\varphi _l-\frac{2\pi j\left( s-l \right)}{d} \right] -\sin \left[ \varphi _{s}^{\prime}-\varphi _{l}^{\prime}-\frac{2\pi j\left( s-l \right)}{d} \right] \right\} \sin \left( \theta _s-\theta _l \right) \right\} 
\\
&=\sum_{s<l \in V}C^{(j)}_{sl} \cos \left( \theta _s-\theta _l \right)+ S^{(j)}_{sl} \sin \left( \theta _s-\theta _l \right), 
\label{eq:deltaQ=0F}
\end{aligned}
\end{equation}
where $$
C^{(j)}_{sl}=\frac{2}{d}{r_sr_l\left\{ \cos \left[ \varphi _s-\varphi _l-\frac{2\pi j\left( s-l \right)}{d} \right] -\cos \left[ \varphi _{s}^{\prime}-\varphi _{l}^{\prime}-\frac{2\pi j\left( s-l \right)}{d} \right] \right\}},$$
$$S^{(j)}_{sl}=\frac{2}{d}{r_sr_l\left\{ \sin \left[ \varphi _s-\varphi _l-\frac{2\pi j\left( s-l \right)}{d} \right] -\sin \left[ \varphi _{s}^{\prime}-\varphi _{l}^{\prime}-\frac{2\pi j\left( s-l \right)}{d} \right] \right\}}.$$




Given the definitions and properties of algebraic numbers, we know that algebraic numbers remain algebraic under addition, subtraction, multiplication, and division by another algebraic number. Therefore, for the states $|\psi\rangle$ and $|\psi'\rangle$ in the set $\mathbb{P}_A$, the following holds true:
(1) The amplitudes $r_k$ and $r_k'$ are algebraic numbers.
(2) The trigonometric functions of the phases, $\cos \varphi_k$, $\sin \varphi_k$, $\cos \varphi_k'$, and $\sin \varphi_k'$ are   algebraic numbers.
(3)  $\cos \left[ \frac{2\pi j\left( s-l \right)}{d} \right] , \sin \left[ \frac{2\pi j\left( s-l \right)}{d} \right]$ are  algebraic numbers where $j,s,l,d$ are integers.
As such, it is easy to prove $C^{(j)}_{sl}$ and $S^{(j)}_{sl}$ are algebraic numbers.

Consider the conditions we mentioned in the main text: 
 
\begin{enumerate}
    \item A.1,  $\theta_l-\theta_k$ is an algebraic number  for $0\le l<k\le d-1$. 
    
    \item A.2, $|\theta_l -\theta_k| $ is not equal to $|\theta_{l'} -\theta_{k'}| $ for any different pair $(l,k)$ and $(l',k')$, where $0\le l<k\le d-1$ and $0\le l'<k'\le d-1$.  
\end{enumerate}
 
If the above two conditions hold, we can deduce that $\{\cos(\theta_l-\theta_k),\sin(\theta_l-\theta_k)\}$ are linearly independent over the field of all algebraic numbers by \autoref{lemma:linear_indep}.  That is, if $C^{(j)}_{sl}$ and $S^{(j)}_{sl}$ are not all 0, there must be 
$$
\varDelta Q_j= \sum_{s<l \in V}C^{(j)}_{sl} \cos \left( \theta _s-\theta _l \right)+ S^{(j)}_{sl} \sin \left( \theta _s-\theta _l \right) \ne 0.
$$
For the case where $C^{(j)}_{sl}$ and $S^{(j)}_{sl}$ are all 0, one has $\varDelta Q_j=0$.  
We shall see that this results in $\varphi_i=\varphi^{\prime}_i$, i.e. $\ket{\psi} = \ket{\psi^\prime}$.  This is because  all $C^{(j)}_{sl}=0$ and $S^{(j)}_{sl}=0$ suggest 
$$
 \cos \left[ \varphi _s-\varphi _l-\frac{2\pi j\left( s-l \right)}{d} \right] -\cos \left[ \varphi _{s}^{\prime}-\varphi _{l}^{\prime}-\frac{2\pi j\left( s-l \right)}{d} \right] =0,$$
$$\sin \left[ \varphi _s-\varphi _l-\frac{2\pi j\left( s-l \right)}{d} \right] -\sin \left[ \varphi _{s}^{\prime}-\varphi _{l}^{\prime}-\frac{2\pi j\left( s-l \right)}{d} \right] =0.$$

There exists $s \in V$. Let us choose to set  $s=g$ and we have $\varphi_s=\varphi_g=0=\varphi^\prime_s=\varphi^\prime_g$. The above equations become
\begin{equation}
     \cos \left[ -\varphi _l-\frac{2\pi j\left( g-l \right)}{d} \right]  =\cos \left[ -\varphi _{l}^{\prime}-\frac{2\pi j\left( g-l \right)}{d} \right], \quad
\sin \left[ -\varphi _l-\frac{2\pi j\left( g-l \right)}{d} \right] =\sin \left[ -\varphi _{l}^{\prime}-\frac{2\pi j\left( g-l \right)}{d} \right].
\label{eq0}
\end{equation}
Since $\varphi_s,\varphi^\prime_s\in[0,2\pi)$, the solution for Eq. \ref{eq0} is $\varphi_s=\varphi^\prime_s$. This analysis hold for any $s\in V$.
Therefore, there is no quantum state $\ket{\psi^\prime} \in \mathbb{P}_A$ can simulate the measurement outcome $\mathbf{P}$ and $\mathbf{Q}$ of a state $\ket{\psi} \in \mathbb{P}_A$  given $\ket{\psi}\ne \ket{\psi^\prime} $.

In other words, satisfying the conditions A.1 and A.2, our two-measurement-bases scheme is informationally complete with respect to all algebraic pure states.
A simple example satisfying the above relationship is
$ 
\alpha _k=2^{k-1}z_0
$
where $z_0$ is an arbitrary algebraic number for all $k$. Another example is
$\theta_k=\sqrt{\alpha_k}$ where $\{\alpha_k:k=0,\cdots,d-1\}$ are positive prime numbers.

\section{Proof of the uniqueness for qubit systems}\label{qubitcase}
\subsection{Derivation of probability distributions for multi-qubit states}
 
Similar to the qudit case, a general $n$-qubit pure state can be written as
\begin{equation}
\begin{aligned}
|\psi \rangle 
=\sum_{k=0}^{2^n-1}{r_ke^{i\varphi _k}}|k\rangle,
\label{state}
\end{aligned}
\end{equation}
where $\sum_{k=0}^{d-1}{r_{k}^{2}=1}$ and  $r_k\geqslant 0$. 
The coefficient $r_k$ can be determined  the probability distribution of the computational basis (the first measurement base in our scheme).
The unitary of the second measurement base of our scheme is given as

\begin{equation}
\begin{aligned}
S^{\prime\dagger}= \bigotimes_{j=1}^n S_{j}^{\dagger} =(\bigotimes_{j=1}^n H^{\dagger} ) D^\dagger =H^{\prime\dagger}D^\dagger .
\end{aligned}
\label{HR}
\end{equation}
where $D^{\dagger}=\sum_{m=0}^{d-1}{e^{-i\theta _m}|m\rangle \langle m|}$,
\begin{equation}
\begin{aligned}
H =\frac{1}{\sqrt{2}}\left[ \begin{matrix}	1&		1\\	1&		-1\\\end{matrix} \right],  \text{and}  \quad H^{\prime}=\bigotimes_{j=1}^n{H}. 
\end{aligned}
\end{equation}


Here we first derive the quantum state after acting $n$-qubit Hadamard gates.
For a single-qubit case, 
$H^{\dagger}|j\rangle =\frac{1}{\sqrt{2}}\left( |0\rangle +\left( -1 \right) ^j|1\rangle \right) =\frac{1}{\sqrt{2}}\sum_{p=0}^{1}{\left( -1 \right) ^{jp}|p\rangle}$. 
Extending to $n$ qubits case, we have
\begin{equation}
\begin{aligned}
H^{\prime\dagger}|j\rangle =\left( H^{\dagger} \right) ^{\otimes n}|j\rangle &=\frac{1}{\sqrt{2^n}}\left[ |0\rangle +\left( -1 \right) ^{j_n}|1\rangle \right] \otimes \left[ |0\rangle +\left( -1 \right) ^{j_{n-1}}|1\rangle \right] \otimes \cdots \otimes \left[ |0\rangle +\left( -1 \right) ^{j_1}|1\rangle \right] \\&=\frac{1}{\sqrt{2^n}}\sum_{p_1,p_2,\cdots,p_n=0}^1{\left( -1 \right) ^{\sum_{m=1}^n{j_mp_m}}|p_n p_{n-1}\cdots p_1\rangle} \\&=\frac{1}{\sqrt{2^n}}\sum_{p_1,p_2,\cdots,p_n=0}^1{\left( -1 \right) ^{\sum_{m=1}^n{j_m\delta _{p_m,1}}}|p_n p_{n-1}\cdots p_1\rangle},
\label{eq:SIColumn}
\end{aligned}
\end{equation}
where $|j\rangle =|j_n j_{n-1}\cdots j_1\rangle , j_1,j_2\cdots ,j_n\in \left\{ 0,1 \right\} $. 
Apply to the general form of quantum state (Eq. \ref{state}) we get
\begin{equation}
\begin{aligned}
|\psi ^H\rangle =H^{\prime\dagger}|\psi \rangle =\sum_{j=0}^{2^n-1}{\psi _{j}^{H}|j\rangle},
\end{aligned}
\end{equation}
where $\psi _{j}^{H}=\frac{1}{\sqrt{2^n}}\sum_{k=0}^{2^n-1}{\left( -1 \right) ^{j_1\delta _{k_1,1}+j_2\delta _{k_2,1}+\cdots +j_n\delta _{k_n,1}}r_ke^{i\varphi _k}}, |j\rangle =|j_n j_{n-1}\cdots j_1\rangle , |k\rangle =|k_n k_{n-1}\cdots k_1\rangle $.
The square of the $j$-th amplitude of the state now becomes
\begin{equation}
\begin{aligned}
Q_{j}(H^{\prime\dagger}\ket{\psi})=\left| \langle j|\psi ^H\rangle \right|^2=\psi _{j}^{H}\psi _{j}^{H\dagger}=\frac{1}{2^n}+\frac{1}{2^{n-1}}\sum_{s<l}{\left( -1 \right) ^{h\left( j,s \right) +h\left( j,l \right)}r_sr_l\cos \left( \varphi _s-\varphi _l \right)},
\label{eq:SIProb}
\end{aligned}
\end{equation}
while $h\left( j,s \right) =j_1\delta _{s_1,1}+j_2\delta _{s_2,1}+\cdots +j_n\delta _{s_n,1}$, $h\left( j,l \right) =j_1\delta _{l_1,1}+j_2\delta _{l_2,1}+\cdots +j_n\delta _{l_n,1}$, $|s\rangle =|s_n s_{n-1}\cdots s_1\rangle, |l\rangle =|l_{n}l_{n-1}\cdots l_1\rangle $.

Now the probability distribution of our second measurement base is given as
\begin{equation}
\begin{aligned}
Q_j\left( S^{\prime\dagger}|\psi \rangle \right) =Q_{j}\left( H^{\prime\dagger}|\phi \rangle \right) &=\frac{1}{2^n}+\frac{1}{2^{n-1}}\sum_{s<l}{\left( -1 \right) ^{h\left( j,s \right) +h\left( j,l \right)}r_sr_l\cos \left( \varphi _s- \theta_{s}-\varphi _l+\theta_{l} \right)}\\&=\frac{1}{2^n}+\frac{1}{2^{n-1}}\sum_{s<l}{\left( -1 \right) ^{h\left( j,s \right) +h\left( j,l \right)}r_sr_l\cos \left[ \left( \varphi _s-\varphi _l \right) -\left( \theta _{s}-\theta_{l}\right) \right]},
\label{eq:Prob2}
\end{aligned}
\end{equation}
where $
|\phi \rangle 
=
D^\dagger\ket{\psi} 
=\sum_{k=0}^{2^n-1}{e^{-i\theta _{k}}|k\rangle \langle k|\psi \rangle}=\sum_{j=0}^{2^n-1}{r_je^{i\left( \varphi _j-\theta_{j} \right)}|j\rangle}$.
It can be seen that the measurement results of the second basis are equivalent to shifting the phases of the quantum state from $\varphi _j$ to $\varphi _j-\theta _{j}$ in the \autoref{eq:SIProb}.

\subsection{ Proof of uniqueness of multi-qubit states $\in \mathbb{P}_A$}


Suppose that there exists such an algebraic state $|\psi ^{\prime}\rangle \in \mathbb{P}_A$ can simulate the measurement outcomes $\mathbf{P}$ and $\mathbf{Q}$ of proposed two measurement bases of $\ket{\psi}\in \mathbb{P}_A$.
Here we assume $|\psi \rangle=\sum_{j=0}^{2^n-1}{r_je^{i\varphi _j}}|j\rangle$ and set $P_j = r_j^2$ denoting the $j$th component of $\mathbf{P}$ and $Q_j=Q_j\left( S^{\prime\dagger} |\psi \rangle \right)$ (see Eq. \ref{eq:Prob2}) denoting the $j$th component of $\mathbf{Q}$. 

Let us analyze the conditions that $|\psi' \rangle \in \mathbb{P}_A$ should satisfy. First $|\psi ^{\prime}\rangle $ can simulate the unknown algebraic state $|\psi \rangle $ under the computational base, e.g.
 $P_j^{\prime}=P_j$.
Therefore, without losing generality, we suppose $|\psi \rangle =\sum_{j=0}^{2^n-1}{r_je^{i\varphi _j}|j\rangle}$ and $|\psi ^{\prime}\rangle =\sum_{j=0}^{2^n-1}{r_je^{i\varphi _{j}^{\prime}}|j\rangle}$, where $r_j\geqslant 0$, $\varphi ,\varphi ^{\prime}\in \left[ \left. 0,2\pi \right) \right. $ ( $r_j$, $r^\prime_j$, $\cos(\phi_j)$, $\cos(\phi^\prime_j)$,   $\sin(\phi_j)$ and $\sin(\phi^\prime_j)$ are algebraic numbers).
Given that the global phase has no physical observation effect, we first choose $g$ from the set $V=\{k\left|r_k\ne 0 \right. \}$, where $k$ is an integer that varies from $0$ to $2^n-1$ and simultaneously satisfies $r_k \ne 0$. Here, $g$ represents the smallest ordinal number of the non-zero amplitudes.
We then set its corresponding phase $\varphi_g = 0$ and $\varphi_g' = 0$, which means that we use it as the reference standard to define the remaining phases. It is important to note that if the amplitude of a component is zero, the value of its phase is meaningless.
Therefore a more accurate description of the unknown quantum state should be 
\begin{equation}
\begin{aligned}
|\psi \rangle =\sum_{j\in V}{r_je^{i\varphi _j}|j\rangle}=\sum_{j\in V,\, j\ne g}{r_je^{i\varphi _j}|j\rangle}+r_g|g\rangle.
\label{eq:fai1F}
\end{aligned}
\end{equation}
The corresponding quantum state $|\psi ^{\prime}\rangle $ that simulates the measurement result of unknown quantum state $|\psi \rangle $ under the computational base is
\begin{equation}
\begin{aligned}
|\psi ^{\prime}\rangle =\sum_{j\in V}{r_je^{i\varphi^\prime_j}|j\rangle}=\sum_{j\in V,\, j\ne g}{r_je^{i\varphi _{j}^{\prime}}|j\rangle}+r_g|g\rangle. 
\label{eq:fai2F}
\end{aligned}
\end{equation}
Since $|\psi ^{\prime}\rangle $ is assume to be a state different from $|\psi \rangle $, there is at least one different phase with nonzero amplitude between $\ket{\psi}$ and $\ket{\psi^\prime}$.

Let analyse whether $|\psi ^{\prime}\rangle $ can simulate the unknown algebraic state $|\psi \rangle $ under the second basis. If it is true, we have $\varDelta Q_j=Q_j-Q_{j}^{\prime}=0 $, where $Q_j^\prime
$ donates the measurement outcome of the simulated state $\ket{\psi
^\prime}$. 
Now we calculate the difference of the probability, e.g.
\begin{equation}
\begin{aligned}
\varDelta Q_j=Q_j-Q_{j}^{\prime}=\frac{1}{2^{n-1}}\sum_{s<l\in V}{\left( -1 \right) ^{h\left( j,s \right) +h\left( j,l \right)}r_sr_l\left\{ \cos \left[ \left( \varphi _s-\varphi _l \right) -\left( \theta_s-\theta_l \right) \right] -\cos \left[ \left( \varphi _{s}^{\prime}-\varphi _{l}^{\prime} \right) -\left( \theta_s-\theta_l \right) \right] \right\}}.
\label{eq:SIparadox}
\end{aligned}
\end{equation}
Expanding $\varDelta Q_j$ yields:
\begin{equation}
\begin{aligned}
\varDelta Q_j &=\frac{1}{2^{n-1}}\sum_{s<l\in V}{\left( -1 \right) ^{h\left( j,s \right) +h\left( j,l \right)}r_sr_l\left\{ \left[ \cos \left( \varphi _s-\varphi _l \right) -\cos \left( \varphi _{s}^{\prime}-\varphi _{l}^{\prime} \right) \right] \cos \left( \theta_s-\theta_l \right) +\left[ \sin \left( \varphi _s-\varphi _l \right) -\sin \left( \varphi _{s}^{\prime}-\varphi _{l}^{\prime} \right) \right] \sin \left( \theta_s-\theta_l \right) \right\}}
\\
&=\sum_{s<l \in V}C^{(j)}_{sl} \cos \left( \theta _s-\theta _l \right)+ S^{(j)}_{sl} \sin \left( \theta _s-\theta _l \right), 
\label{eq:SIdeltaQ=0}
\end{aligned}
\end{equation}
where
\begin{equation}
\begin{aligned}
&C^{(j)}_{sl} =\frac{1}{2^{n-1}}{\left( -1 \right) ^{h\left( j,s \right) +h\left( j,l \right)}r_sr_l\left[ \cos \left( \varphi _s-\varphi _l \right) -\cos \left( \varphi _{s}^{\prime}-\varphi _{l}^{\prime} \right) \right]},\\
&S^{(j)}_{sl}=\frac{1}{2^{n-1}}{\left( -1 \right) ^{h\left( j,s \right) +h\left( j,l \right)}r_sr_l\left[ \sin \left( \varphi _s-\varphi _l \right) -\sin \left( \varphi _{s}^{\prime}-\varphi _{l}^{\prime} \right) \right]}.
\label{C1S1}
\end{aligned}
\end{equation}
Note that (1) The amplitudes $r_k$ and $r_k'$ are algebraic numbers.
(2) The trigonometric functions of the phases, $\cos \varphi_k$, $\sin \varphi_k$, $\cos \varphi_k'$, and $\sin \varphi_k'$ are   algebraic numbers. Similar to the qudit case, 
it is easy to prove $C^{(j)}_{sl}$ and $S^{(j)}_{sl}$ are algebraic numbers.
Consider the following conditions: 
 
\begin{enumerate}
    \item A.1,  $\theta_l-\theta_k$ is an algebraic number  for $0\le l<k\le 2^n-1$. 
    
    \item A.2, $|\theta_l -\theta_k| $ is not equal to $|\theta_{l'} -\theta_{k'}| $ for any different pair $(l,k)$ and $(l',k')$, where $0\le l<k\le 2^n-1$ and $0\le l'<k'\le 2^n-1$.  
    \label{A1A2}
\end{enumerate}
If the above two conditions hold, we can deduce that $\{\cos(\theta_l-\theta_k),\sin(\theta_l-\theta_k)\}$ are linearly independent over the field of all algebraic numbers by \autoref{lemma:linear_indep}.  That is, if $C^{(j)}_{sl}$ and $S^{(j)}_{sl}$ are not all 0, there must be 
$$
\varDelta Q_j= \sum_{s<l \in V}C^{(j)}_{sl} \cos \left( \theta _s-\theta _l \right)+ S^{(j)}_{sl} \sin \left( \theta _s-\theta _l \right) \ne 0.
$$
For the case where $C^{(j)}_{sl}$ and $S^{(j)}_{sl}$ are all 0, one has $\varDelta Q_j=0$.  
We shall see that this results in $\varphi_i=\varphi^{\prime}_i$, i.e. $\ket{\psi} = \ket{\psi^\prime}$.  This is because  all $C^{(j)}_{sl}=0$ and $S^{(j)}_{sl}=0$ suggest 
$$
 \cos \left( \varphi _s-\varphi _l \right) -\cos \left( \varphi _{s}^{\prime}-\varphi _{l}^{\prime}\right) =0,$$
$$\sin \left( \varphi _s-\varphi _l \right) -\sin \left( \varphi _{s}^{\prime}-\varphi _{l}^{\prime}\right) =0.$$

There exist $s \in V$. Let us choose to set  $s=g$ and we have $\varphi_s=\varphi_g=0=\varphi^\prime_s=\varphi^\prime_g$. The above equations become
\begin{equation}
     \cos \left( -\varphi _l\right)  =\cos \left( -\varphi _{l}^{\prime}\right), \quad
\sin \left( -\varphi _l\right) =\sin \left( -\varphi _{l}^{\prime} \right).
\label{eq0}
\end{equation}
Since $\varphi_s,\varphi^\prime_s\in[0,2\pi)$, the solution for Eq. \ref{eq0} is $\varphi_s=\varphi^\prime_s$. This analysis holds for any $s\in V$.
Therefore, there is no quantum state $\ket{\psi^\prime} \in \mathbb{P}_A$ can simulate the measurement outcome $\mathbf{P}$ and $\mathbf{Q}$ of a state $\ket{\psi} \in \mathbb{P}_A$  given $\ket{\psi}\ne \ket{\psi^\prime} $.

In other words, satisfying the conditions A.1 and A.2, our two-measurement-bases scheme is informationally complete with respect to all algebraic pure states.
A simple example satisfying the above relationship is
$ 
\alpha _k=2^{k-1}z_0
$
where $z_0$ is an arbitrary algebraic number for all $k$. Another example is
$\theta_k=\sqrt{\alpha_k}$ where $\{\alpha_k:k=0,\cdots,d-1\}$ are positive prime numbers.

\subsection{Uniqueness for Sparse state $\in \mathbb{P}_A$}

Our above proof shows for any pure algebraic state, two measurement bases are informationally complete. 
However, as the state dimension increases, the complexity of implementing the second measurement basis also grows. For an $n$-qubit system, where $d=2^n$, the implementation of a global phase gate becomes exponentially complex in the general case. 

Given the measurement distribution of the first basis, the implementation of the second basis can be significantly simplified. This is because applying phase operations to components that are zero in the computational basis is unnecessary. Consequently, if the number of nonzero components $r_k$ of the quantum state in the computational basis scales polynomially with $n$, i.e. $O(poly(n))$, the operation $D^\dagger$ for the second basis can be defined as follows:
\begin{equation}
 D^{\dagger}=\sum_{m=0}^{d-1}{e^{-i\theta _m(1-\delta_{r_m,0})}|m\rangle \langle m|} = \sum_{m\in V} e^{-i\theta_m} \ket{m}\bra{m}+\sum_{m\ne V} \ket{m}\bra{m}
\end{equation}
where $\delta_{r_m,0}$ is the delta function and $V$ is defined as the set with a non-zero component of the quantum state (see \autoref{eq:fai1F}). This indicates that $D^\dagger$ can be realised by polynomial elementary gates.

One can verify that the uniqueness of the sparse state $\in \mathbb{P_A}$ with simplified implementation also holds since the proof is the same as the general one.


\section{Discussions the uniqueness of a set of $\mathbb{P}_A$ for two-local measurement bases}\label{local}
For the local measurement bases case, 
one may use $\bigotimes_{j=1}^n{R^{\dagger}\left( \alpha _j \right)}$ to replace the entangled diagonal phase gate $D^\dagger$, i.e. 
\begin{equation}
\begin{aligned}
(D^\prime)^\dagger=\bigotimes_{j=1}^n{R^{\dagger}\left( \alpha _j \right)}=\sum_{m=0}^{2^n-1}{e^{-i\alpha _{m}^{\prime}}|m\rangle \langle m|},
\label{phase}
\end{aligned}
\end{equation}
where $|m\rangle =|m_nm_{n-1}\cdots m_1\rangle$ and $\alpha _{m}^{\prime}=\sum_{p=1}^n{\alpha _pm_p}=\sum_{p=1}^n{\alpha _p\delta _{m_p,1}}$. 
Here $R(\alpha_j)=R_j$ is defined as follows,
\begin{equation}
R_j=\begin{bmatrix}
1 & 0 \\
0 & e^{i\alpha_j} \nonumber\\
\end{bmatrix}.
\end{equation}
Thus, the unitary of the second bases is equivalent to acting a diagonal phase gate $\bigotimes_{j=1}^n{R^{\dagger}\left( \alpha _j \right)}=\sum_{m=0}^{2^n-1}{e^{-i\alpha _{m}^{\prime}}|m\rangle \langle m|}$ on the quantum state and then applying the Hadamard gate on each qubit. 

Similar to the analysis in \autoref{qubitcase}, expanding $\varDelta Q_j$ yields:
\begin{equation}
\begin{aligned}
\varDelta Q_j &=\frac{1}{2^{n-1}}\sum_{s<l\in V}{\left( -1 \right) ^{h\left( j,s \right) +h\left( j,l \right)}r_sr_l\left\{ \left[ \cos \left( \varphi _s-\varphi _l \right) -\cos \left( \varphi _{s}^{\prime}-\varphi _{l}^{\prime} \right) \right] \cos \left( \alpha _{s}^{\prime}-\alpha _{l}^{\prime} \right) +\left[ \sin \left( \varphi _s-\varphi _l \right) -\sin \left( \varphi _{s}^{\prime}-\varphi _{l}^{\prime} \right) \right] \sin \left( \alpha _{s}^{\prime}-\alpha _{l}^{\prime} \right) \right\}}
\\
&=\sum_{ \varDelta _t }{f_1\left( j|\varDelta _t \right) \cos \varDelta _t+f_2\left( j|\varDelta _t \right) \sin \varDelta _t},
\label{eq:SIdeltaQ=0}
\end{aligned}
\end{equation}
where $\varDelta _t=\alpha _{s}^{\prime}-\alpha _{l}^{\prime}$ represents $t$-th different relative phases (for local bases, there may exist many relative phases of $\{\alpha_s^\prime\}$ that are the same), and we have:
\begin{equation}
\begin{aligned}
f_1\left( j|\varDelta _t \right) =\frac{1}{2^{n-1}}\sum_{s<l\in V}{\left( -1 \right) ^{h\left( j,s \right) +h\left( j,l \right)}r_sr_l\left[ \cos \left( \varphi _s-\varphi _l \right) -\cos \left( \varphi _{s}^{\prime}-\varphi _{l}^{\prime} \right) \right]}_{\varDelta _t=\alpha _{s}^{\prime}-\alpha _{l}^{\prime}},\\
f_2\left( j|\varDelta _t \right) =\frac{1}{2^{n-1}}\sum_{s<l\in V}{\left( -1 \right) ^{h\left( j,s \right) +h\left( j,l \right)}r_sr_l\left[ \sin \left( \varphi _s-\varphi _l \right) -\sin \left( \varphi _{s}^{\prime}-\varphi _{l}^{\prime} \right) \right]}_{\varDelta _t=\alpha _{s}^{\prime}-\alpha _{l}^{\prime}}.
\label{f1f2}
\end{aligned}
\end{equation}
Here, we essentially merge terms that have the same relative phase. Note that there is a difference between this arrangement and the qudit case. In the qudit case, the measurement phase of the second basis can be arbitrary. However, in the case of local measurements, the phase corresponding to the measurement basis is a partial or full summation of the individual local phases.

For our local basis, $\alpha _{m}^{\prime}-\alpha _{m^{\prime}}^{\prime}=\sum_{p=1}^n{\alpha _pm_p}-\sum_{q=1}^n{\alpha _qm^\prime_q}$ (see Eq. \ref{phase}).  Generally,  any relative phase can be expressed as the following general formula:
\begin{equation}
  \sum_{m\in M}(-1)^{x_m} \alpha_m,
\end{equation}
where $M$ is the subset of the set $N=\{1,2,3...,n\}$. 
The restrictions of $\{\alpha_m\}$ is that
given  $\Vec{x}=\{x_m\}$  is not equal to  $\Vec{x^\prime}=\{x_m^\prime\}$,
\begin{equation}
\begin{aligned}
\left|\sum_{m\in M}(-1)^{x_m} \alpha_m \right|\ne \left|\sum_{m\in M}(-1)^{x_m^\prime} \alpha_m \right|,
\end{aligned}
\label{eq_theta_qubits_SM}
\end{equation}
where all $\alpha_m$ are algebraic number. That is $\Delta_t  \ne \Delta_{t^\prime}$ if $t\ne t^{\prime}$. 
Note that $\ket{\psi},\ket{\psi^\prime }\in \mathbb{P}_A$,  $f_1\left( j|\varDelta _t \right)$ and $f_2\left( j|\varDelta _t \right)$ are algebraic numbers or 0.
 We can deduce that $\{\cos \Delta_t,\sin \Delta_t \}$ are linearly independent over the field of all algebraic numbers by \autoref{lemma:linear_indep}.  That is, if $f_1\left( j|\varDelta _t \right)$ and $f_2\left( j|\varDelta _t \right)$ are not all 0, there must be 
$$
\varDelta Q_j= \sum_{ \varDelta _t }{f_1\left( j|\varDelta _t \right) \cos \varDelta _t+f_2\left( j|\varDelta _t \right) \sin \varDelta _t}\ne 0.
$$

For the case where $f_1\left( j|\varDelta _t \right)$ and $f_2\left( j|\varDelta _t \right)$ are all 0, one has $\varDelta Q_j=0$. 
In this case, the situation is complicated, and whether $\varphi_j=\varphi^{\prime}_j$ with $j\in V$ is an open problem.
Thus, satisfying the above restriction of $\{\alpha_i\}$, our two-local measurement scheme is informationally complete with respect to a set of algebraic pure states (satisfying $f_1\left( j|\varDelta _t \right)$ and $f_2\left( j|\varDelta _t \right)$ are not all 0).

Similar to the qudit case, a simple example satisfying the above constraints is
$ 
\alpha _k=2^{k-1}z_0
$
 for any $k$, where $z_0$ is an arbitrary algebraic number.
Another example is $
\alpha _k=\sqrt{q_k}
$
where $q_k$ is the $k$th prime number sorted from smallest to largest.

\subsection{Information Completeness: W-like States and GHZ-like States}

Here we demonstrate that two local bases are sufficient to characterise two special classes of sparse algebraic  states: W-like algebraic 
 states and GHZ-like algebraic states.

A general W-like algebraic state can be defined as 
\begin{equation}
    |\text{W-like}\rangle=\sum_{j=1}^n r_j e^{i\varphi_j} |0\rangle^{\otimes (j-1)} \otimes |1\rangle \otimes |0\rangle^{\otimes (n-j)}.
\end{equation}

The W-like state has $n$ amplitudes that are nonzero. After performing local phase gate $\bigotimes_{j=1}^{n}R^\dagger_j$ ($R^\dagger_j=\ket{0}\bra{0}+e^{-\alpha_j}\ket{1}\bra{1}$), the state becomes
\begin{equation}
    \bigotimes_{j=1}^{n}R^\dagger_j|\text{W-like}\rangle=\sum_{j=1}^n r_j e^{i(\varphi_j-\alpha_j)} |0\rangle^{\otimes (j-1)} \otimes |1\rangle \otimes |0\rangle^{\otimes (n-j)}.
\end{equation}
One can see that this phase-shifted state has the same form as the state after the global phase gate $D^\dagger=\sum_{m=1}^{d-1} e^{-i\theta_m} \ket{m}\bra{m}$, i.e. $D^\dagger|\psi_{\text{W}}\rangle=\sum_{j=1}^n r_j e^{i(\varphi_j-\theta_j)} |0\rangle^{\otimes (j-1)} \otimes |1\rangle \otimes |0\rangle^{\otimes (n-j)}$. Since $R_j$ is local we may set $\alpha_j$ independently. Setting $\alpha_l$ is algebraic number and $|\alpha_l-\alpha_s|\ne |\alpha_i-\alpha_j|$ for any $l,s\ne ij$, the \autoref{f1f2} becomes

\begin{equation}
\begin{aligned}
f_1\left( j|\varDelta _t \right) =\frac{1}{2^{n-1}}{\left( -1 \right) ^{h\left( j,s \right) +h\left( j,l \right)}r_sr_l\left[ \cos \left( \varphi _s-\varphi _l \right) -\cos \left( \varphi _{s}^{\prime}-\varphi _{l}^{\prime} \right) \right]},\\
f_2\left( j|\varDelta _t \right) =\frac{1}{2^{n-1}}{\left( -1 \right) ^{h\left( j,s \right) +h\left( j,l \right)}r_sr_l\left[ \sin \left( \varphi _s-\varphi _l \right) -\sin \left( \varphi _{s}^{\prime}-\varphi _{l}^{\prime} \right) \right]},
\end{aligned}
\end{equation}
which is exactly same as \autoref{C1S1}.  Thus, the uniqueness is guaranteed by the linear independence theorem \autoref{lemma:linear_indep}.


Similarly, GHZ-like algebraic  states defined as 
\begin{equation}
|\text{GHZ-like}\rangle = \frac{1}{\sqrt{2}} \left( \sum_{j=1}^n r_0 |0\rangle^{\otimes n} + r_1 e^{i\varphi} |1\rangle^{\otimes n} \right),
\end{equation}

and its phase-shifted state 
\begin{equation}
 \bigotimes_{j=1}^{n}R^\dagger_j|\text{GHZ-like}\rangle = \sum_{j=1}^n r_0 |0\rangle^{\otimes n} + r_1 e^{i(\varphi-\sum_j\alpha_j)} |1\rangle^{\otimes n} .
\end{equation}
GHZ-like states also satisfy the uniqueness of two local bases. We note that the second local basis may be further simplified to $R_1\otimes I^{\otimes n-1}$. This is because a GHZ-like state only has two components and a single local phase gate is enough to shift the phase of the second component.

Moreover, we can use a similar technique to find other sparse algebraic states that can be determined by two local bases. We need to consider the nonzero terms such that  the phase-shifting 
for all nonzero amplitudes $r_l$ and $r_k$ satisfy the conditions A1 and A.2.

\section{Proof of \autoref{thm:many_solution}}


\begin{proof}
      
    It has been rigorously demonstrated that at least four measurement bases are required to uniquely determine a pure state, so if only two measurement bases are used, there exists a group of pure states which can not be uniquely determined. 
    
    We prove the second statement by an example.
    Consider the $n$-qubit GHZ-like state as the unknown target state
    \begin{equation}
    |\psi \rangle =a_0|0\rangle ^{\otimes n}+a_{2^n-1}e^{i\varphi _{2^n-1}}|1\rangle ^{\otimes n}.
    \end{equation}
    Consider the $n$-qubit simulated state as
    \begin{equation}
    |\psi ^{\prime}\rangle =\sum_{j=0}^{2^n-1}{a_{j}^{\prime}e^{i\varphi _{j}^{\prime}}|j\rangle},
    \end{equation}
    and set the global phase $\varphi _{0}^{\prime}=0$.
    From the measurement results of the first measurement base we can get
    \begin{equation}
    a_{0}^{\prime}=a_0, a_{2^n-1}^{\prime}=a_{2^n-1}, a_{l}^{\prime}=a_l=0, l=1,2,\cdots ,2^n-2.
    \end{equation}
    That is, the structure of the simulated state at this point becomes
    \begin{equation}
    |\psi ^{\prime}\rangle =a_0|0\rangle ^{\otimes n}+a_{2^n-1}e^{i\varphi _{2^n-1}^{\prime}}|1\rangle ^{\otimes n}.
    \end{equation}
    Next, by taking the measurements of the second measurement base and calculating $\varDelta Q_j=Q_j-Q_{j}^{\prime}$, we can get
    \begin{equation}
    \cos \left( \varphi _{2^n-1}-\alpha _{2^n-1}^{\prime} \right) -\cos \left( \varphi _{2^n-1}^{\prime}-\alpha _{2^n-1}^{\prime} \right) =0.
    \end{equation}
    At this point it can be solved that
    \begin{equation}
    \varphi _{2^n-1}^{\prime}=\varphi _{2^n-1}\,\,or\,\,\varphi _{2^n-1}^{\prime}=-\varphi _{2^n-1}+2\alpha _{2^n-1}^{\prime}.
    \end{equation}
    These two solutions may be far apart. It is easy to verify  the inner product of these solutions is $a_0^2+a^2_{2^n-1}e^{i(2\varphi _{2^n-1}-2\alpha _{2^n-1}^{\prime} )}$, which can be small.

    However, if we assume that both the target state and the simulated state are algebraic states, then it is natural to obtain the unique solution $\varphi _{2^n-1}^{\prime}=\varphi _{2^n-1}$.
    This is because another solution $\varphi _{2^n-1}^{\prime}=-\varphi _{2^n-1}+2\alpha _{2^n-1}^{\prime}$ will let
    \begin{equation}
    \cos \left( \varphi _{2^n-1}^{\prime} \right) =\cos \left( -\varphi _{2^n-1}+2\alpha _{2^n-1}^{\prime} \right) =\cos \left( \varphi _{2^n-1} \right) \cos \left( 2\alpha _{2^n-1}^{\prime} \right) +\sin \left( \varphi _{2^n-1} \right) \sin \left( 2\alpha _{2^n-1}^{\prime} \right), 
    \end{equation}
    \begin{equation}
    \sin \left( \varphi _{2^n-1}^{\prime} \right) =\sin \left( -\varphi _{2^n-1}+2\alpha _{2^n-1}^{\prime} \right) =-\sin \left( \varphi _{2^n-1} \right) \cos \left( 2\alpha _{2^n-1}^{\prime} \right) +\cos \left( \varphi _{2^n-1} \right) \sin \left( 2\alpha _{2^n-1}^{\prime} \right). 
    \end{equation}
    This suggest the simulated state is not an algebraic state, which in turn is excluded.\\
    
\end{proof}



While \autoref{thm:many_solution} seems a bit negative, it indeed is concerned with a different objective of tomography. The existence of multiple solutions is because this is a question of asking if there exist states that can re-produce the probability distribution.
However, universal quantum computation only requires the implementation of circuits such as Clifford + T circuits, or circuits consisting of single-qubit rotating gates (the rotation angle should be $\beta \pi$, where $\beta$ is a rational number) + CNOT gates, \sun{which are the building blocks of  broad applications in quantum chemistry simulation.} These circuits still make the evolving quantum state in $\mathbb{P}_A$. In these scenarios our scheme is valid.

\section{Extension to mixed states with known types of noise}

Our two-bases scheme remains valid for mixed states with known type of noise.
The measurement and reconstruction process is shown in \autoref{eq:mix1} and \autoref{eq:mix2}.
We measure it under two measurement bases to obtain two sets of probability distributions $\mathbf{P},\,\,\mathbf{Q}$, where $P_j=\rho _{jj},\,\,Q_j=\rho _{jj}^{\prime}$, while the mixed state is later reconstructed by post-processing.
The difference from the case of pure state is that the target quantum state is added with known type of noise, so in the reconstruction process we can set the form of simulated quantum state with known type of noise.
\begin{equation}
\begin{aligned}
I\rho I=\rho \xrightarrow{measurement}\mathbf{P}
\\
S^{\prime\dagger}\rho S^{\prime}=\rho ^{\prime}\xrightarrow{measurement}\mathbf{Q}
\label{eq:mix1}
\end{aligned}
\end{equation}
\begin{equation}
\begin{aligned}
\left. \begin{array}{r}
	\mathbf{P}\\
	\mathbf{Q}\\
\end{array} \right\} \xrightarrow{reconstruction}\rho 
\label{eq:mix2}
\end{aligned}
\end{equation}

\hspace*{\fill}

\textbf{(I) Depolarisation Noise}

First, consider the common Werner-like state which involves depolarising noise.
It means that we already know both the target state and simulated state take the following form
\begin{equation}
\begin{aligned}
\rho =p|\psi \rangle \langle \psi |+\left( 1-p \right) \frac{I}{d}.
\end{aligned}
\end{equation}

Suppose that the depolarisation coefficient $p$ is a nonzero algebraic number but the exact magnitude is unknown.
Like the contradiction analysis for a pure state, first assuming that there exists a simulated mixed state $\rho ^{\prime}=p^{\prime}|\psi ^{\prime}\rangle \langle \psi ^{\prime}|+\left( 1-p^{\prime} \right) \frac{I}{d}$ different from the true mixed state $\rho =p|\psi \rangle \langle \psi |+\left( 1-p \right) \frac{I}{d}$, the same probability distributions of measurements $\mathbf{P},\,\,\mathbf{Q}$ can be obtained under the $I,S^{\prime\dagger}$ measurement operations.
Note that $|\psi \rangle =\sum_j{r_je^{i\varphi _j}|j\rangle}, |\psi ^{\prime}\rangle =\sum_j{r_{j}^{\prime}e^{i\varphi _{j}^{\prime}}|j\rangle}, d=2^n$.
First the simulated mixed state $\rho ^{\prime}$ can simulate the unknown mixed state $\rho$ under the computational base.
There should be $P_{j}^{\prime}=P_j$, and thus we could get
\begin{equation}
\begin{aligned}
p\left( r_j \right) ^2=P_j-\frac{1-p}{d},\,\,\,\, p^{\prime}\left( r_{j}^{\prime} \right) ^2=P_j-\frac{1-p^{\prime}}{d}.
\label{eq:rj}
\end{aligned}
\end{equation}

Next $\rho ^{\prime}$ can simulate the unknown mixed state $\rho$ under the second basis.
From the results under the second measurement base, we can analogize the result of the pure state in the \autoref{eq:Prob2} and get the difference of the probability distributions
\begin{equation}
\begin{aligned}
\varDelta Q_j=\varDelta \rho _{jj}^{\prime}&=Q_j-Q_{j}^{\prime}
\\
&=\frac{1}{2^{n-1}}\sum_{s<l}{\left( -1 \right) ^{h\left( j,s \right) +h\left( j,l \right)}\left\{ pr_sr_l\cos \left[ \left( \varphi _s-\varphi _l \right) -\left( \alpha _{s}^{\prime}-\alpha _{l}^{\prime} \right) \right] -p^{\prime}r_{s}^{\prime}r_{l}^{\prime}\cos \left[ \left( \varphi _{s}^{\prime}-\varphi _{l}^{\prime} \right) -\left( \alpha _{s}^{\prime}-\alpha _{l}^{\prime} \right) \right] \right\}} =0,
\label{eq:mixedpro}
\end{aligned}
\end{equation}
and we notice that the summation term in \autoref{eq:mixedpro} can be expanded as
\begin{footnotesize}
 \begin{equation}
 \begin{aligned}
&pr_sr_l\cos \left[ \left( \varphi _s-\varphi _l \right) -\left( \alpha _{s}^{\prime}-\alpha _{l}^{\prime} \right) \right] -p^{\prime}r_{s}^{\prime}r_{l}^{\prime}\cos \left[ \left( \varphi _{s}^{\prime}-\varphi _{l}^{\prime} \right) -\left( \alpha _{s}^{\prime}-\alpha _{l}^{\prime} \right) \right] 
\\
=&pr_sr_l\cos \left( \varphi _s-\varphi _l \right) \cos \left( \alpha _{s}^{\prime}-\alpha _{l}^{\prime} \right) +pr_sr_l\sin \left( \varphi _s-\varphi _l \right) \sin \left( \alpha _{s}^{\prime}-\alpha _{l}^{\prime} \right) -p^{\prime}r_{s}^{\prime}r_{l}^{\prime}\cos \left( \varphi _{s}^{\prime}-\varphi _{l}^{\prime} \right) \cos \left( \alpha _{s}^{\prime}-\alpha _{l}^{\prime} \right) -p^{\prime}r_{s}^{\prime}r_{l}^{\prime}\sin \left( \varphi _{s}^{\prime}-\varphi _{l}^{\prime} \right) \sin \left( \alpha _{s}^{\prime}-\alpha _{l}^{\prime} \right) 
\\
=&\left[ pr_sr_l\cos \left( \varphi _s-\varphi _l \right) -p^{\prime}r_{s}^{\prime}r_{l}^{\prime}\cos \left( \varphi _{s}^{\prime}-\varphi _{l}^{\prime} \right) \right] \cos \left( \alpha _{s}^{\prime}-\alpha _{l}^{\prime} \right) +\left[ pr_sr_l\sin \left( \varphi _s-\varphi _l \right) -p^{\prime}r_{s}^{\prime}r_{l}^{\prime}\sin \left( \varphi _{s}^{\prime}-\varphi _{l}^{\prime} \right) \right] \sin \left( \alpha _{s}^{\prime}-\alpha _{l}^{\prime} \right). 
\label{eq:mixedpro2}
 \end{aligned}
 \end{equation}
\end{footnotesize}

If the exact value of depolarising noise $p$ is known, which means that $p^{\prime}=p$, then the amplitude of the quantum state can be uniquely determined through \autoref{eq:rj}.
Meanwhile, the noise coefficient in \autoref{eq:mixedpro} can be eliminated and thus the equation becomes the same form as \autoref{eq:SIparadox}.
Same as our analysis of the pure state, we get $|\psi ^{\prime}\rangle =|\psi \rangle$. 
Overall if it is known that $p^{\prime}=p$, we must have $\rho ^{\prime} = \rho$ in this case. 
 
\textbf{(II) Noise for Any Known Noise Coefficient and Noise Type}

Consider a general form of noise  
\begin{equation}
\begin{aligned}
\rho =p|\psi \rangle \langle \psi |+\left( 1-p \right) \sigma, 
\end{aligned}
\end{equation}
where $\sigma$ may be a thermal state or other mixed state is known a priori. If the exact magnitude of the noise coefficient and the noise type are known, then the scenario is same to our analysis for the case of \textbf{(I)}, where the amplitude of the quantum state can be uniquely determined and the difference of the probability distributions under the second measurement base becomes the same form as \autoref{eq:SIparadox}.
In this case our scheme can uniquely determine the mixed state.
 
\textbf{(III) Noise robustness}

In practical situations, it is unknown whether the quantum state contains a small amount of noise, i.e., the target state is generally a mixed state of high purity. And it is difficult to analyze this situation by modeling the probability distribution of a mixed state in the form of a pure state.
In the next section, we will verify that our scheme has a degree of stability and noise tolerance through numerical simulations.

\section{Numerical Results}

\subsection{Numerical setup}

Let us re-express the quantum state as 
\begin{equation}
|\psi \rangle =\sum_{k=0}^{2^n-1}{r_ke^{i\varphi _k}|k\rangle}=\sum_{k=0}^{2^n-1}{\left( a_k+ib_k \right) |k\rangle}.
\label{eq:psi_a_b}
\end{equation}
The relationships between $r_k,\varphi _k$ and $a_k,b_k$ are
\begin{equation}
\begin{aligned}
r_{k}^{2}=a_{k}^{2}+b_{k}^{2},\,\, \cos \varphi _k=\frac{a_k}{\sqrt{a_{k}^{2}+b_{k}^{2}}},\,\, \sin \varphi _k=\frac{b_k}{\sqrt{a_{k}^{2}+b_{k}^{2}}}. 
\label{eq:one}
\end{aligned}
\end{equation}

Three conditions need to be satisfied in the ideal two-bases scheme:

\textbf{($1$)} Both target states and simulated states are algebraic states, i.e., the state components $a_k, b_k, a_{k}^{\prime}, b_{k}^{\prime}$ of $|\psi \rangle =\sum_{k=0}^{2^n-1}{\left( a_k+ib_k \right)}|k\rangle$ and $|\psi^{\prime}\rangle =\sum_{k=0}^{2^n-1}{\left( a_{k}^{\prime}+ib_{k}^{\prime} \right)}|k\rangle$ are \textbf{algebraic numbers}.

\textbf{($2$)} All the $e^{i\alpha}$ in the single-qubit gates
\begin{equation}
\begin{aligned}
T=\left[ \begin{matrix}
	1&		0\\
	0&		e^{i\alpha}\\
\end{matrix} \right].
\end{aligned}
\end{equation}
of the second basis are \textbf{transcendental numbers}. We note that this transcendental operation is achievable in physical implementation but not classical computer.
All the $\alpha$ we choose are square root of different prime numbers.

\textbf{($3$)} The values of the probability distribution measured under the first basis are algebraic numbers, and the values of the probability distribution measured under the second basis are mostly \textbf{transcendental numbers}.

\begin{figure}[t]
    \centering
\includegraphics[width=0.45\textwidth]{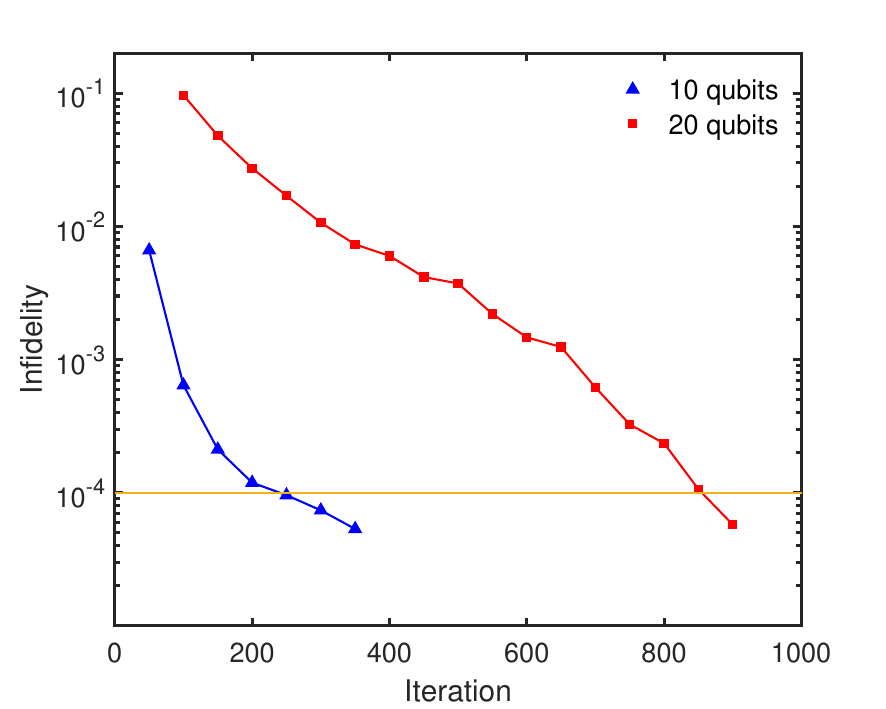}

    \caption{ \textbf{10-qubit and 20-qubit W-like states reconstruction.} 
    The blue triangular dots represent changes in fidelity during the reconstruction process of the 10-qubit W-like state. The red square dots represent changes in fidelity during the reconstruction process of the 20-qubit W-like state. Infidelity: $1-$Fidelity. The orange line: fidelity over $99.99\%$.
    The algebraic numbers and transcendental numbers are simulated on a computer using low-precision numbers and high-precision numbers, respectively, as computers can only store data in finite decimal places. 
    The numbers of decimal places for the modulus $r_k$, the phases $\cos \varphi_k$ and  $\sin \varphi_k$ are denoted as $C_1$, $C_2$, and $C_3$, respectively, and are set to be $3$.
    The numbers of decimal places for the second measurement base are $X=15$, and for the measured probability distributions are $Y=15$. The target state $\ket{\psi}$ and the simulated state $\ket{\psi'}$ are set at the same level of precision.
    }
    \label{fig:M1_Iteration_Results}
\end{figure}

\begin{figure}[t]
    \centering
    \includegraphics[width=\textwidth]{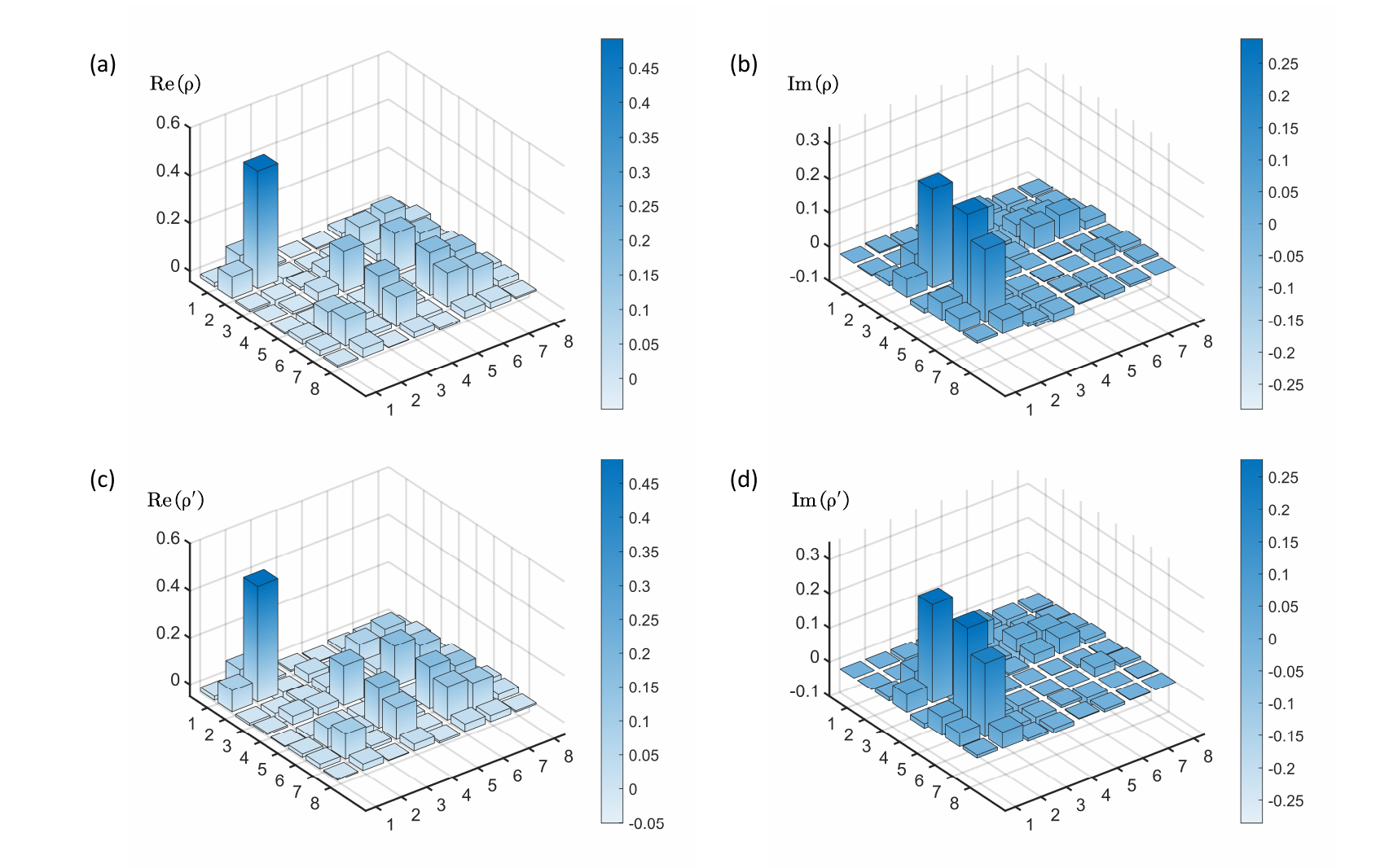}
    \caption{The set-up is the same as that in Fig.~2 in the main text. The numbers of decimal places for the amplitudes are $C_1=C_{1}^{\prime}=3$, for the phases are $C_2=C_{2}^{\prime}=3$, and we set $C_2=C_3\,\,\left( C_{2}^{\prime}=C_{3}^{\prime} \right)$. The numbers of decimal places for the second measurement base are $X=X^{\prime}=15$, and for the measured probability distributions are $Y=Y^{\prime}=15$. (a)(b) Real and imaginary parts of the target quantum state. (c)(d) Real and imaginary parts of the reconstructed quantum state. }
    \label{SI1_8_dimensional_state}
\end{figure}

\begin{figure}[t]
    \centering
    \includegraphics[width=\textwidth]{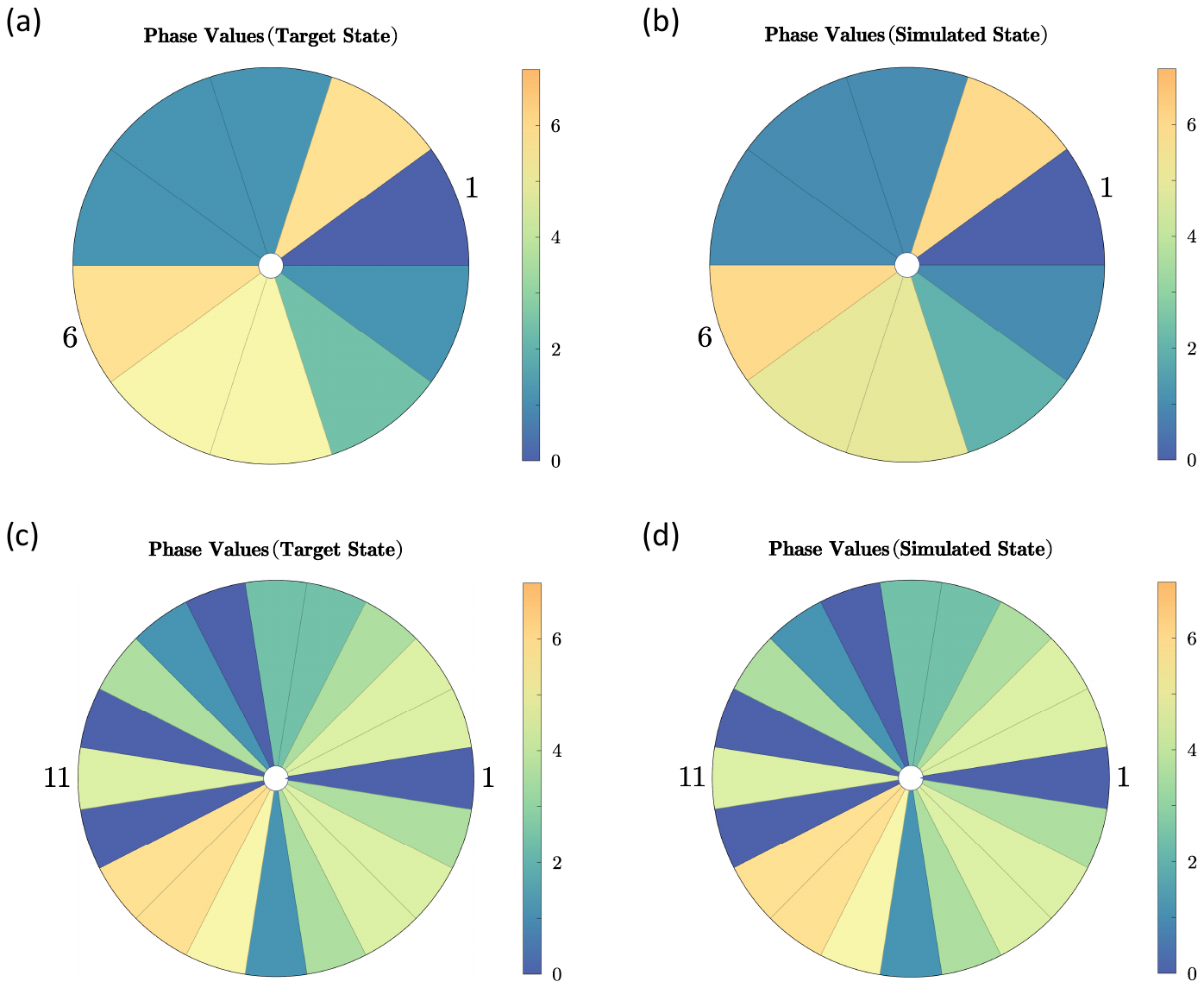}
    \caption{10-qubit and 20-qubit W-like states reconstruction results. \textbf{a,b}, Phase values of the target state and the reconstructed simulated state in the form of a 10-qubit W-like state. \textbf{c,d}, Phase values of the target state and the reconstructed simulated state in the form of a 20-qubit W-like state. (a)(b)(c)(d) The phase values of the corresponding components from smallest to largest in the subscripts of the nonzero components of the quantum states are shown in order along the counter-clockwise direction starting from the number $1$ in the figure. }
    \label{SI2_10_and_20_qubits}
\end{figure}

In these cases, the reconstruction of the target state can be realized with two bases.
However, for the three requirements mentioned above, the second and third points cannot be fulfilled on existing classical computers.
In fact, computers can only store up to a finite number of digits after the decimal point (we set the number of digits as $X$). 
For a 64-bit computer, the number of decimal places with respect to the floating-point number is 15 \cite{8766229}, which is widely used in software for algorithm design and simulation.
We can also sacrifice computing speed to get a larger $X$ by programming the calculation process on classical computers.
Based on this, we propose an experimental-friendly two-bases scheme that can be easily implemented on a classical computer. The above three conditions are simplified into the following:

\textbf{($1^{\prime}$)} $a_k, b_k$ are numbers with $M_{\mathbf{1}}$ decimal places, while $a_{k}^{\prime}, b_{k}^{\prime}$ are numbers with $M_{\mathbf{2}}$ decimal places. Equally important is the fact that $a_k, b_k$ and $a_{k}^{\prime}, b_{k}^{\prime}$ both satisfy amplitude normalisation and phase normalisation with $N_{\mathbf{1}}$, $N_{\mathbf{2}}$ and $N_{\mathbf{3}}$, $N_{\mathbf{4}}$ decimal places respectively, which means that
\begin{equation}
\begin{aligned}
\left| \sum_k{r_{k}^{2}}-1 \right|<\frac{1}{2}\times 10^{-N_1},\,\,\left| \cos ^2\varphi _k+\sin ^2\varphi _k-1 \right|<\frac{1}{2}\times 10^{-N_2},
\label{eq:52}
\end{aligned}
\end{equation}
and
\begin{equation}
\begin{aligned}
\left| \sum_k{\left( r_{k}^{\prime} \right) ^2}-1 \right|<\frac{1}{2}\times 10^{-N_3},\,\,\left| \cos ^2\varphi _{k}^{\prime}+\sin ^2\varphi _{k}^{\prime}-1 \right|<\frac{1}{2}\times 10^{-N_4}.
\label{eq:53}
\end{aligned}
\end{equation}
The conversion between $a_k, b_k, a_{k}^{\prime}, b_{k}^{\prime}$ and $r_k, \varphi _k, r_{k}^{\prime}, \varphi _{k}^{\prime}$ is given by \autoref{eq:one}.

\textbf{($2^{\prime}$)} All the $e^{i\alpha}$ in the single-qubit gates of the second basis are numbers with $X$ decimal places in target states, and $X^{\prime}$ decimal places in simulated states (Both $X$ and $X^{\prime}$ are greater than $M_{\mathbf{1}}$ and $M_{\mathbf{2}}$).

\textbf{($3^{\prime}$)} The numbers in probability distributions of target states $\mathbf{P},\,\,\mathbf{Q}$ are numbers with $Y$ decimal places, and probability distributions of simulated states $\mathbf{P^{\prime}},\,\,\mathbf{Q^{\prime}}$ with $Y^{\prime}$ decimal places (Both $Y$ and $Y^{\prime}$ are greater than $M_{\mathbf{1}}$ and $M_{\mathbf{2}}$).

\subsection{Numerical simulations}

We randomly generate an eight-dimensional algebraic state as our target state $|\psi \rangle$. 
The real and imaginary parts of the density matrix of the target state are shown in Fig.~\ref{SI1_8_dimensional_state}(a)(b).
The real and imaginary parts of the density matrix of the reconstructed state are shown in Fig.~\ref{SI1_8_dimensional_state}(c)(d).
The fidelity of the reconstructed state is $99.9998\%$ and can further approach $100\%$ by increasing the number of iterations.

Then, we scale up the simulation by considering the reconstruction with a large problem size of up to 20 qubits.
To enable the numerical simulation, we choose to set the unknown quantum state in the form of an n-qubit W-like state defined above.
The aim is to recover the phase information from the probability distribution under two measurement bases.
The randomly selected phase values of the target state in the form of a 10-qubit W-like state and the phase values of its reconstructed state are shown in Fig.~\ref{SI2_10_and_20_qubits}(a) and Fig.~\ref{SI2_10_and_20_qubits}(b), respectively.
Meanwhile, the randomly selected phase values of the target state in the form of a 20-qubit W-like state and the phase values of its reconstructed state are shown in Fig.~\ref{SI2_10_and_20_qubits}(c) and Fig.~\ref{SI2_10_and_20_qubits}(d), respectively.
The fidelity changes of the reconstructed states during the iteration process for the 10-qubit W-like state and 20-qubit W-like state are shown in Fig.~\ref{fig:M1_Iteration_Results} in the main text.
The fidelity of the reconstructed states can all reach more than $99.99\%$.

\end{document}